\documentclass{article}

\usepackage[preprint]{neurips_2026}
\usepackage{amssymb}
\usepackage{amsmath}
\usepackage{multirow}
\usepackage{wrapfig}
\usepackage{caption}
\usepackage{sidecap}
\usepackage{placeins}
\PassOptionsToPackage{numbers, compress}{natbib}
\workshoptitle{Interpretability for Discovery}

\usepackage[utf8]{inputenc} 
\usepackage[T1]{fontenc}    
\usepackage{hyperref}       
\usepackage{url}            
\usepackage{booktabs}       
\usepackage{amsfonts}       
\usepackage{nicefrac}       
\usepackage{microtype}      
\usepackage{xcolor}         

\usepackage{graphicx}
\usepackage{booktabs}
\usepackage{array}

\newcommand{\gridimg}[1]{%
  \raisebox{-.5\height}{%
    \includegraphics[width=0.19\textwidth]{#1}
  }%
}

\newcommand{\pairlabel}[1]{%
  \raisebox{-0.35ex}{\scriptsize\bfseries Pair #1}%
}

\newcommand{\threepairgrid}[1]{%
\begingroup
\setlength{\tabcolsep}{1.2pt}
\renewcommand{\arraystretch}{0.98}

\begin{tabular}{
    @{}r
    @{\hspace{4pt}}
    c
    @{\hspace{2pt}}
    c
    @{\hspace{14pt}}
    c
    @{\hspace{2pt}}
    c
    @{}
}

&
\multicolumn{2}{c}{\normalsize\bfseries Flagged}
&
\multicolumn{2}{c}{\normalsize\bfseries Matched control}
\\[-0.10em]

&
{\footnotesize Legacy}
&
{\footnotesize HSC}
&
{\footnotesize Legacy}
&
{\footnotesize HSC}
\\[0.20em]

\pairlabel{1}
&
\gridimg{Figures/adaptive_cp/#1_pair1_legacy_flagged.pdf}
&
\gridimg{Figures/adaptive_cp/#1_pair1_hsc_flagged.pdf}
&
\gridimg{Figures/adaptive_cp/#1_pair1_legacy_control.pdf}
&
\gridimg{Figures/adaptive_cp/#1_pair1_hsc_control.pdf}
\\[0.05em]

\pairlabel{2}
&
\gridimg{Figures/adaptive_cp/#1_pair2_legacy_flagged.pdf}
&
\gridimg{Figures/adaptive_cp/#1_pair2_hsc_flagged.pdf}
&
\gridimg{Figures/adaptive_cp/#1_pair2_legacy_control.pdf}
&
\gridimg{Figures/adaptive_cp/#1_pair2_hsc_control.pdf}
\\[0.05em]

\pairlabel{3}
&
\gridimg{Figures/adaptive_cp/#1_pair3_legacy_flagged.pdf}
&
\gridimg{Figures/adaptive_cp/#1_pair3_hsc_flagged.pdf}
&
\gridimg{Figures/adaptive_cp/#1_pair3_legacy_control.pdf}
&
\gridimg{Figures/adaptive_cp/#1_pair3_hsc_control.pdf}

\end{tabular}
\endgroup
}

\title{Do Astronomical Foundation Models \\Know When They Will Fail?}

\author{
\textbf{Vaidehi Bulusu}$^{1,2,3}$
\qquad
\textbf{Mehul Goyal}$^{2,3,4}$
\qquad
\textbf{Md Rabius Sany Apu}$^{2,3}$
\\[0.25em]
\textbf{Michael J. Smith}$^{2,3,5,6}$
\qquad
\textbf{Shashwat Sourav}$^{2,3,7}$
\\[0.45em]
\small
$^{1}$Columbia University
\quad
$^{2}$EleutherAI
\quad
$^{3}$UniverseTBD
\\[-0.05em]
\small
$^{4}$Indian Institute of Science Education and Research, Thiruvananthapuram
\\[-0.05em]
\small
$^{5}$AstroAI
\quad
$^{6}$Center for Astrophysics | Harvard \& Smithsonian
\quad
$^{7}$Washington University in St. Louis
}

\begin{document}

\maketitle

\begin{abstract}
Astronomical foundation models are increasingly used as reusable representations across surveys, but it is unclear when the same object remains consistently represented across instruments or whether a model can warn us before transfer fails. We present a conformal framework for auditing cross-survey representations using cross-matched Legacy Survey--HSC, HSC--JWST, and SDSS--HSC objects. Astronomy-specific AstroPT embeddings show lower translation errors and tighter conformal regions than general-purpose visual backbones at comparable coverage, while different model families fail on different objects. We then use adaptive conformal prediction to test whether the source embedding contains information about its own future error. Using a ridge difficulty model fitted to source embeddings, we find that this warning signal is consistent across the tested backbones for Legacy$\rightarrow$HSC, backbone-dependent for HSC$\rightarrow$JWST, and largely absent for SDSS$\rightarrow$HSC. When informative, adaptation improves coverage for the hardest objects. Matched-control analysis further reveals silent failures not explained by magnitude, apparent size, or nearest-neighbor distance. Hence, we show that conformal uncertainty provides a realistic way to identify when astronomical representations transfer reliably and when they fail without warning.
\end{abstract}

\section{Introduction}


Astronomical foundation models are increasingly being used as reusable representations of survey data \citep{2024arXiv240514930S,2024arXiv241202527T}. But the same galaxy can look very different across instruments because surveys differ in depth, angular resolution, wavelength coverage, modalities and observing conditions \citep{2019AJ....157..168D,2018PASJ...70S...4A,2018ConPh..59..251K}. This raises a question that downstream benchmarks do not answer: if an embedding is useful in one survey, can we trust it to represent the same object consistently in another? We study this question using cross-matched Legacy Survey-HSC, HSC-JWST, and SDSS-HSC objects. Our main finding is that cross-survey reliability is itself a property of the learned representation. We find that some foundation models transfer much more precisely than others, different model families fail on different galaxies, and some representations contain a measurable warning that their cross-survey transfer is about to fail, while others fail silently. We use conformal prediction \citep{vovk2005algorithmic} to turn these failures into calibrated uncertainty regions and then inspect the galaxies for where the uncertainty succeeds or breaks. Large astronomical foundation models such as AstroPT have shown that self-supervised pretraining can produce useful representations from millions of galaxy images \citep{2024arXiv240514930S}, while datasets such as the Multimodal Universe make large-scale learning across astronomical surveys and modalities possible \citep{2024arXiv241202527T}. Conformal prediction has also begun to be applied in astronomy. Recent work studies uncertainty for radio-galaxy classification \citep{2026arXiv260320000W} and for physical-property prediction from astronomical foundation-model embeddings \citep{2026arXiv260607771T}. These works calibrate uncertainty around a downstream prediction. We instead use conformal prediction to audit the \emph{representation itself}. We view this as a representation-level interpretability problem. Beyond measuring transfer error, the goal is to identify what the learned embedding reveals about when and where that transfer will fail.

\begin{figure*}[htbp]
    \centering
    \includegraphics[width=\linewidth]{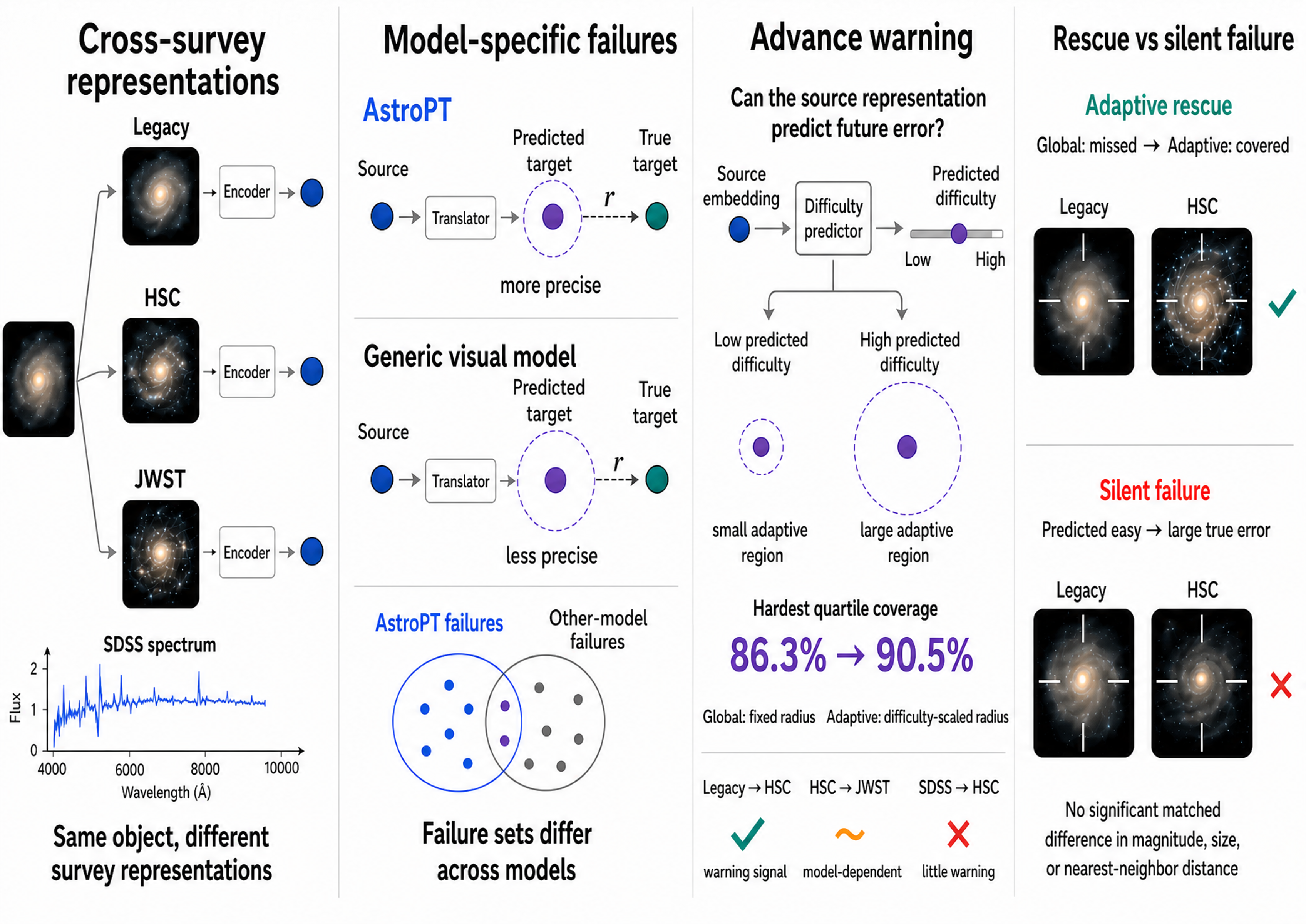}
    \caption{\textbf{Cross-survey representation audit.} \textbf{(a)} The same object can have different survey representations. \textbf{(b)} Models differ in transfer precision and failure cases. \textbf{(c)} Adaptive conformal prediction tests whether source embeddings anticipate transfer difficulty. \textbf{(d)} Matched examples distinguish adaptive rescues from silent failures.
}
    \label{fig:overview}
\end{figure*}

\noindent\textbf{Contributions.}
We (i) benchmark cross-survey transfer across astronomical and general-purpose representations; (ii) show that different families fail on different objects
and test whether those failures can be anticipated; and (iii) connect the resulting failure modes to matched galaxy images and survey metadata. Figure~\ref{fig:overview} gives a summary of our analysis. 

and test whether those failures can be anticipated

\section{Methodology}
\label{sec:method}


\textbf{Data and representations.}
We use cross-matched objects from Legacy Survey \citep{2019AJ....157..168D}$\rightarrow$HSC \citep{2018PASJ...70S...4A}, HSC$\rightarrow$JWST \citep{2024arXiv241202527T}, and SDSS \citep{2000AJ....120.1579Y}$\rightarrow$HSC\footnote{Datasets:
\url{https://huggingface.co/astro-conformal-project}}. The first two compare image representations of the same object across imaging surveys, SDSS$\rightarrow$HSC maps a shared SDSS spectral representation \citep{parker2024astroclip} to an HSC image representation. For the imaging surveys we evaluate AstroPT \citep{2024arXiv240514930S}, DINOv2 \citep{2023arXiv230407193O}, ConvNeXt \citep{2022arXiv220103545L}, I-JEPA \citep{2023arXiv230108243A}, and ViT \citep{2020arXiv201011929D}, with several model sizes. We use fixed train, calibration, and test splits throughout, and L2-normalize every embedding before translation.

\textbf{Cross-survey translation.}
For each backbone, we learn a mapping from the source-survey embedding
to the corresponding target-survey embedding. We use ridge regression \citep{10.2307/1271436}, $k$-nearest-neighbor regression \citep{0f828a91-f0ca-3041-a6aa-3ba93ade8ce0}, and radial basis function (RBF) support-vector regression \citep{2998981.2999003} rather than a
large neural translator: the goal is to measure how predictable the target representation already is from the source representation, not to build a
powerful generative model. Agreement across translators also checks that the
result reflects the embedding space rather than one particular regression
method. For object $i$, we measure translation error using the dimension-normalized residual $ r_i =
    \left\|
    \mathbf e^{\,t}_i -
    f(\mathbf e^{\,s}_i)
    \right\|_2
    /\sqrt d$
where $d$ is the target embedding dimension. Smaller residuals mean that the
target-survey representation is more predictable from the source survey. We
report the mean, median, and 90th percentile residuals.

\textbf{Global conformal calibration.} After fitting the translator on the training split, we compute residuals on a held-out calibration split. For target coverage $1-\alpha$, the global radius is $q_{1-\alpha} =
\operatorname{Quantile}\!\left(
\{r_i\}_{i=1}^{n_{\rm cal}},
\frac{\lceil(n_{\rm cal}+1)(1-\alpha)\rceil}{n_{\rm cal}}
\right)$, which gives a marginal coverage guarantee of at least $1-\alpha$ for a new test point using a finite calibration set, under exchangeability \citep{angelopoulos2022gentle}. This radius defines an $L_2$-ball conformal region around the translated target embedding; at matched coverage, smaller radii correspond to more precise transfer.
A test object is covered when its residual is at most $q_{1-\alpha}$. We evaluate $80\%$, $90\%$, and $95\%$ coverage on a separate test split. We compare uncovered-object sets between models with Jaccard overlap \citep{jaccard1901etude},
$J(U_m,U_n)=|U_m\cap U_n|/|U_m\cup U_n|$.

%




\textbf{Adaptive conformal calibration.}
To test whether the source representation predicts its own failure, we train a
separate ridge difficulty model $\hat{\sigma}(x)$ on data disjoint from the translator-training set. Following locally adaptive conformal methods that scale residuals by an estimate of local prediction difficulty
\citep{lei2018distribution,Yun2026.05.28.26354326}, we use calibration scores as $a_i=\frac{r_i}{\hat{\sigma}(x_i)+\epsilon},
\,\,
R_{\rm adapt}(x)=q^a_{1-\alpha}
\bigl(\hat{\sigma}(x)+\epsilon\bigr).$
Thus harder predicted objects receive larger regions. We evaluate the difficulty signal using Spearman correlation with the true residual and coverage across predicted-difficulty quartiles. At $90\%$ coverage we also inspect adaptive rescues, silent failures, and easy-stable objects.

\section{Results}

\subsection{Cross-survey transfer and global conformal prediction}
\label{sec:global_results}

\begin{SCfigure*}[0.7][htbp]
    \centering
    \includegraphics[width=0.8\linewidth]{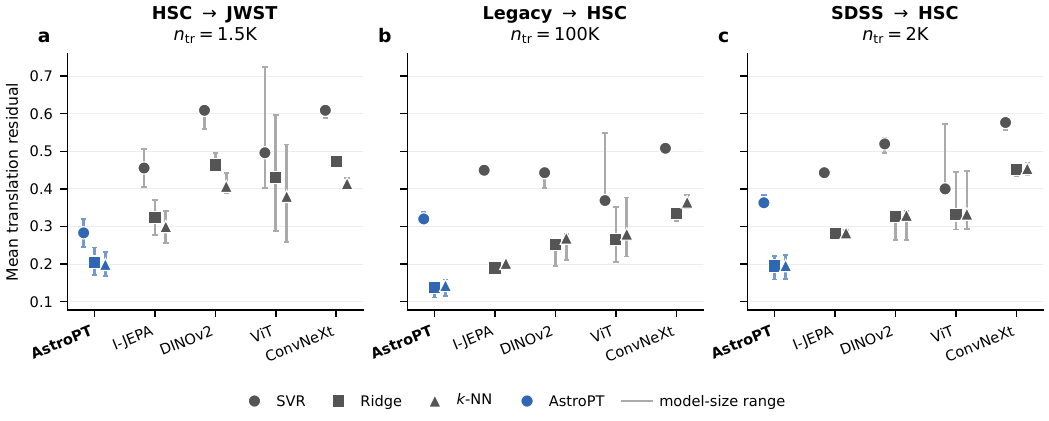}
    \caption{ Cross-survey translation residuals by representation family. The vertical ranges span model sizes. Lower is better; comparisons are within survey pairs.}
    \label{fig:translation_family_summary}
\end{SCfigure*}

\textbf{Translation quality depends strongly on the representation.}
The same cross-survey mapping problem can be substantially easier or harder depending on the embedding space. AstroPT gives the lowest residuals
across the three survey pairs and translator classes
(Figure~\ref{fig:translation_family_summary}). Under ridge translation, AstroPT-850M reaches mean normalized residuals of $0.179$, $0.107$, and $0.168$ for HSC$\rightarrow$JWST, Legacy$\rightarrow$HSC, and SDSS$\rightarrow$HSC, respectively. The ordering is also largely preserved under SVR and $k$-NN translation, suggesting that the difference is a property of the representation rather than a peculiarity of one translator. The full translator comparison is reported in Appendix Table~\ref{tab:residuals-mean}. 

\textbf{Global conformal calibration is close to nominal.}
Before comparing models on their uncertainty regions, we check that the
regions have the expected coverage. Global conformal calibration is close to nominal across all three survey pairs and all three target levels (Appendix Fig.~\ref{fig:global_cp_calibration}).

\begin{figure*}[htbp]
    \centering
    \includegraphics[width=\linewidth]
    {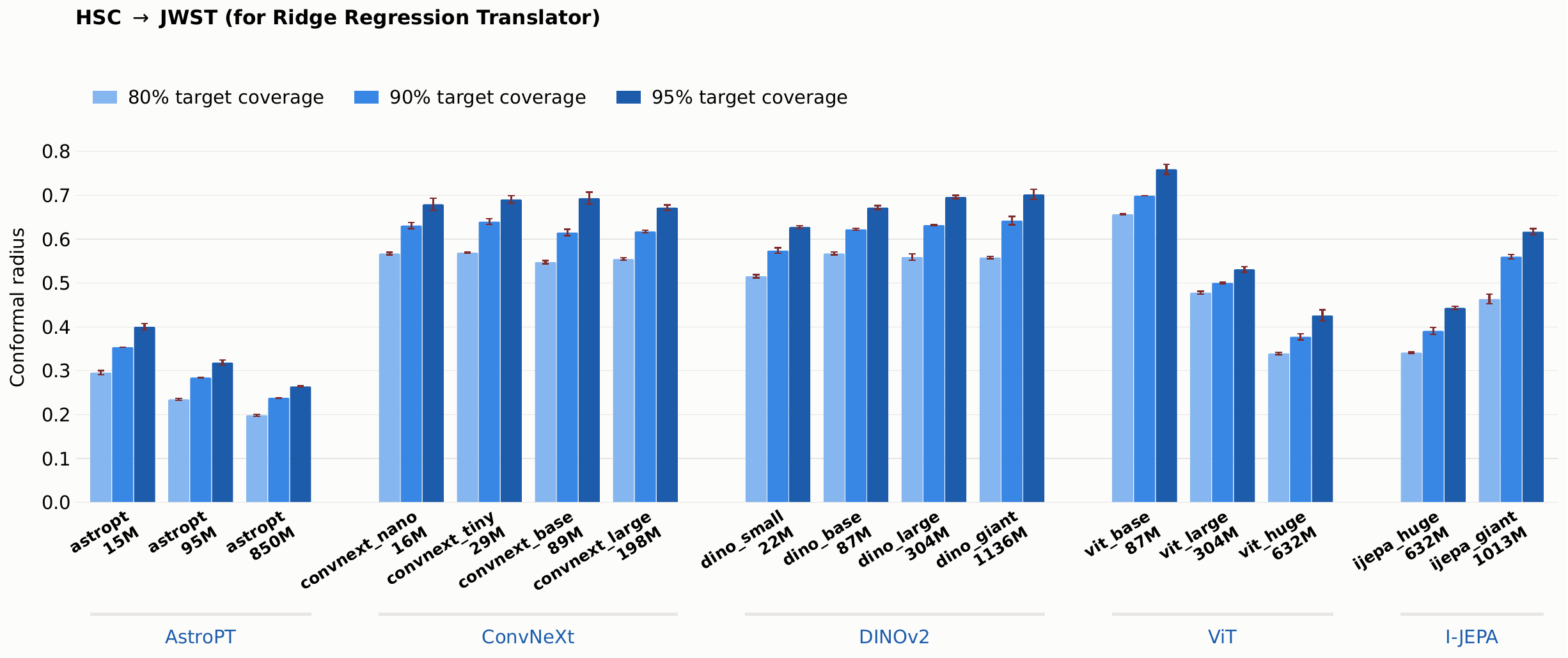}
    \caption{
    Conformal radius required at $80\%$, $90\%$, and $95\%$ target coverage
    for HSC$\rightarrow$JWST translation. Error bars show $\pm1$ standard
    deviation across three calibration/test reshuffles. Smaller regions at
    the same coverage indicate more precise cross-survey transfer.
    }
    \label{fig:global_cp_radius}
\end{figure*}

\textbf{The same coverage can require very different uncertainty regions.}
Figure~\ref{fig:global_cp_radius} shows HSC$\rightarrow$JWST  for the ridge regression translator. AstroPT requires much smaller conformal regions than the generic visual backbones, with AstroPT-850M giving the smallest radius. This means that at the same target coverage, the translated AstroPT embedding can be localized more precisely (within a smaller region) in the JWST embedding space, hence requiring less uncertainty to contain the true target representation. The ordering is stable across repeated calibration/test splits; the full conformal-radius results across all three survey pairs and all three translators are reported in 
Appendix Figures~\ref{fig:global_cp_radius_ridge}--~\ref{fig:global_cp_radius_svr}.


\begin{SCfigure*}[0.7][htbp]
    \centering
    \includegraphics[width=0.7\linewidth]
    {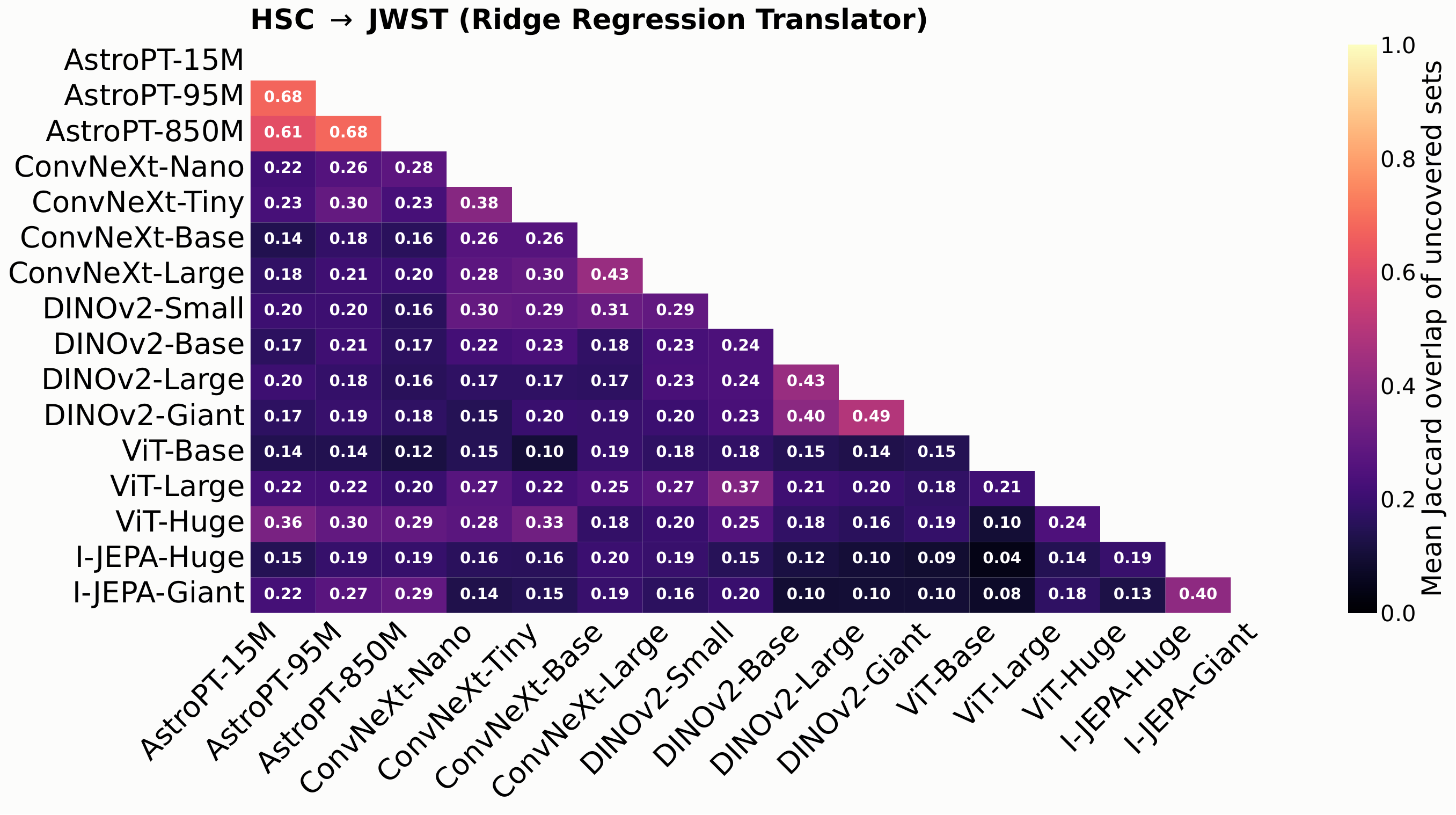}
    \caption{
    Pairwise Jaccard overlap of uncovered-object sets for
    HSC$\rightarrow$JWST at $90\%$ target coverage with ridge regression translator.
    $J=1$ denotes identical failure sets and $J=0$ disjoint sets.
    Models within the same representation family tend to fail on more
    similar objects than models from different families.
    }
    \label{fig:global_cp_overlap}
\end{SCfigure*}

\textbf{Similar failure rates hide different failure populations.}
A fixed coverage level does not mean that representations fail on the same objects. Figure~\ref{fig:global_cp_overlap} compares uncovered sets for HSC$\rightarrow$JWST at $90\%$ coverage with ridge regression translator. AstroPT variants overlap strongly with one another ($J=0.61$--$0.68$), but much less with most other families. DINOv2 and I-JEPA show similar within-family structure. Thus, cross-survey failure depends strongly on the representation even when overall failure rates are similar. Appendix Figures~\ref{fig:ridge_additional_fsilure_overlap}--~\ref{fig:other_additional_fsilure_overlap} extend this analysis to remaining survey pairs and translators.



\subsection{Can models anticipate their own failures?}
\label{sec:adaptive_cp}

\begin{figure*}[htbp]
    \centering
    \includegraphics[width=\linewidth]{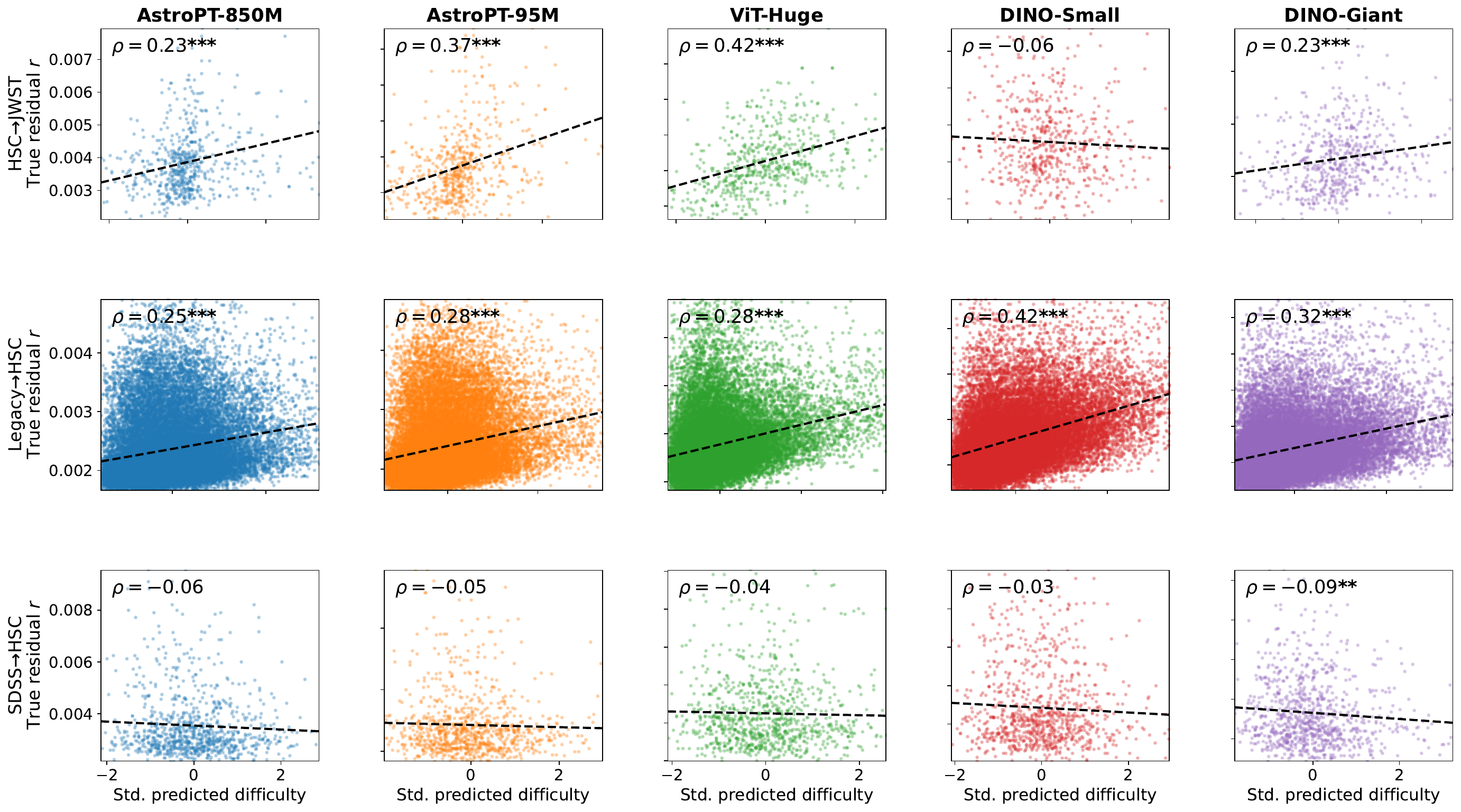}
    \caption{True translation residual versus predicted cross-survey difficulty at $90\%$ target coverage using the ridge difficulty model. Difficulty is standardized separately for each backbone; $\rho$ is the Spearman
    correlation and dashed lines show linear trends. Legacy$\rightarrow$HSC
    contains a consistent warning signal, HSC$\rightarrow$JWST is
    backbone-dependent, and SDSS$\rightarrow$HSC provides little usable
    advance warning.}
    \label{fig:difficulty_residual}
\end{figure*}

\textbf{Failure predictability depends on both the model and survey pair.}
Figure~\ref{fig:difficulty_residual} compares predicted difficulty against the true translation residual at $90\%$ target coverage. Here, predicted difficulty is the residual magnitude expected from the
source-survey embedding alone, before the target-survey representation is
observed. Legacy$\rightarrow$HSC shows a consistent positive relationship across the tested backbones ($\rho=0.25$--$0.42$), indicating that the source representation contains information about its own future translation error. The signal is less uniform for HSC$\rightarrow$JWST: ViT-Huge and AstroPT-95M reach $\rho=0.42$ and $0.37$, respectively, while DINO-Small is essentially uninformative ($\rho=-0.06$). For SDSS$\rightarrow$HSC, the correlations are close to zero across backbones. A warning signal therefore exists for some representation--survey combinations, but is not a generic consequence of fitting a difficulty model.


\textbf{Adaptive calibration improves coverage for objects predicted to be difficult.}
Appendix Figure~\ref{fig:quartile_coverage} groups test objects by predicted difficulty. Global conformal prediction is marginally calibrated, but this can hide substantial conditional undercoverage among objects predicted to be hard. For Legacy$\rightarrow$HSC, the highest-difficulty quartile improves from $86.3\%$ to $90.5\%$ coverage for AstroPT-850M and from $84.3\%$ to $90.4\%$ for AstroPT-95M after adaptation. The same pattern appears for the informative HSC$\rightarrow$JWST backbones: the hardest quartile improves by roughly $13$ percentage points for ViT-Huge and $8$ points for AstroPT-95M.
The SDSS$\rightarrow$HSC setting provides a useful negative control. Because predicted difficulty is largely unrelated to the true residual there, adaptive calibration does not produce a consistent improvement.


\textbf{What distinguishes the objects for which adaptation succeeds or
fails?}
We examine Legacy$\rightarrow$HSC with AstroPT-850M, where the difficulty signal is reliable and the sample is large enough for matched analysis. Adaptive rescues are objects missed by the global region but covered after
adaptation, silent failures have low predicted difficulty despite a large
true residual; easy-stable objects provide the comparison population. The matched analysis gives a more selective picture than the raw image examples. After Holm correction \citep{holm1979simple} across the $12$ comparisons, adaptive-rescue objects differ from easy-stable controls in nearest-neighbor distance ($p_{\mathrm{Holm}}<0.001$), while neither $i$-band magnitude nor apparent size remains significant. Silent failures show no corrected
difference from stable controls in magnitude, apparent size, or nearest-neighbor distance. Their failure is therefore not captured by these simple observables. False alarms, by contrast, differ from correctly anticipated hard objects in nearest-neighbor distance
($p_{\mathrm{Holm}}<0.001$), but not in magnitude or apparent size. Residual differences are reported for completeness, since residual-based outcomes contribute to the category definitions. Representative matched images and the complete statistical comparisons are provided in Appendix Figures~\ref{fig:adaptive_rescue_examples} and Table~\ref{tab:adaptive_effect}. These examples preserve the same flagged control pairing across Legacy Survey and HSC imaging, allowing survey-dependent morphology and local environment to be inspected directly. Finally, we test whether the apparent image differences can be explained only by the higher angular resolution of HSC. We convolve representative HSC cutouts toward the Legacy Survey PSF and repeat the visual comparison (Appendix Figure~\ref{fig:psf_matching}). The degradation suppresses fine structure, but does not uniformly make the flagged HSC objects resemble their Legacy counterparts. Resolution therefore contributes to some cross-survey differences without accounting for the full set of failures.

\section{Discussion and Limitations}
\label{sec:discussion}

Cross-survey reliability depends on the representation, and the astronomical object. Models with similar coverage fail on different galaxies (a failure as an object whose true target-survey embedding falls outside the calibrated conformal region). Some embeddings contain an advance warning of their own transfer error. This makes conformal uncertainty useful as it separates predictable hard cases from failures that occur without warning. Silent failures are particularly interesting because they are not distinguished from stable controls by the simple catalogue quantities we tested.  There are some limitations. First, conformal prediction gives marginal coverage under exchangeability, as it doesn't guarantee the same coverage for every astronomical subpopulation. The weak SDSS$\rightarrow$HSC difficulty signal also shows that object-specific adaptation is not useful in all settings. Second, our interpretation is observational. Matched controls and PSF checks rule out some simple explanations, and do not establish the physical cause of a failure. Larger cross-survey samples and richer physical metadata will be needed to determine whether silent failures form distinct astronomical populations. Thus, our results provide a practical route for using representation failures as targets for scientific inspection.

\section{Acknowledgment}
We would like to thank Stella Biderman and EleutherAI for their compute and infrastructure support. This research used the DeltaAI advanced computing and data resource through allocation number PHY250286, which is supported by the National Science Foundation (award OAC 2320345) and the State of Illinois. DeltaAI is a joint effort of the University of Illinois Urbana-Champaign and its National Center for Supercomputing Applications. Access to DeltaAI was granted through the Advanced Cyberinfrastructure Coordination Ecosystem: Services \& Support (ACCESS) program, which is supported by U.S. National Science Foundation grants \#2138259, \#2138286, \#2138307, \#2137603, and \#2138296.

\bibliographystyle{plainnat}
\bibliography{references}

\clearpage
\appendix

\section{Responsible Use Statement}

This work studies the reliability of existing astronomical foundation-model
representations across surveys. We do not introduce a model intended for
high-stakes decision making or human-facing deployment, and we do not use
sensitive personal data. We expect the main benefit of this work to be improved understanding of when learned representations can be transferred reliably across instruments, which may help researchers identify unreliable model outputs before using them in downstream scientific analyses. 

\section{Reproducibility}
\label{app:repro}

The cross-matched embedding datasets and fixed splits used in this work are available through the anonymous Hugging Face repository linked in the main text. We report the translators, calibration procedure, model backbones, and evaluation metrics required to reproduce the experiments. Full code and model configuration files will be released upon deanonymization.

\section{Additional cross-survey results}
\label{app:additional_global}

\subsection{Translation residuals across backbones and translators}
\label{app:translation_residuals}

Table~\ref{tab:residuals-mean} reports the complete translation benchmark. The relative ordering of the representation families is similar under ridge regression, support-vector regression, and $k$-nearest-neighbor regression. AstroPT gives the lowest residuals in most settings, whereas the generic ConvNeXt representations consistently yield substantially larger cross-survey errors. The agreement across translator classes indicates that these differences are not specific to the choice of regression model.

Table ~\ref{tab:residuals-median} reports the median residuals for translators across backbones, and Table ~\ref{tab:residuals-p90} reports the 90th percentile residuals.

\begin{table}[htbp]
\centering\small\setlength{\tabcolsep}{3.2pt}\renewcommand{\arraystretch}{0.95}
\caption{Cross-survey translation residuals (mean $L_2$ distance) for three survey pairs under three translator classes: support vector regression (RBF kernel), ridge regression, and $k$-NN regression.
Please note that the results here use a different random seeds than those used for the results in section 3.1; consequently, the numerical values differ slightly. Parenthesized values below each mean are standard deviations across test objects; significance markers are from a one-sided permutation test against a random-pairing null ($B=10{,}000$), Holm-corrected within panel. The \texttt{astropt} family gives the lowest residuals in every panel. Residuals are not comparable \emph{across} panels: sample size and source encoder differ by column group. In the HSC $\rightarrow$ JWST and Legacy $\rightarrow$ HSC panels the source and target encoders are matched (e.g.\ \texttt{dino\_large\_hsc} $\rightarrow$ \texttt{dino\_large\_jwst}). In the SDSS $\rightarrow$ HSC panel the source is fixed to \texttt{specformer\_base\_sdss} for all rows, so that column group measures spectral-to-image translation rather than same-backbone alignment. \textbf{Significance:} $^{*}p<0.05$, $^{**}p<0.01$, $^{***}p<0.001$.}
\label{tab:residuals-mean}
\begin{tabular}{l ccc ccc ccc}
\toprule
 & \multicolumn{3}{c}{HSC $\rightarrow$ JWST ($n_{\mathrm{tr}}$=1.5K)} & \multicolumn{3}{c}{Legacy $\rightarrow$ HSC ($n_{\mathrm{tr}}$=100K)} & \multicolumn{3}{c}{SDSS $\rightarrow$ HSC ($n_{\mathrm{tr}}$=2K)} \\
\cmidrule(lr){2-4}\cmidrule(lr){5-7}\cmidrule(lr){8-10}
Backbone & SVR & Ridge & $k$-NN & SVR & Ridge & $k$-NN & SVR & Ridge & $k$-NN \\
\midrule
\texttt{astropt\_15m} & \shortstack{0.321$^{**}$\\{\scriptsize (0.067)}} & \shortstack{0.244$^{**}$\\{\scriptsize (0.078)}} & \shortstack{0.231$^{**}$\\{\scriptsize (0.114)}} & \shortstack{0.338$^{**}$\\{\scriptsize (0.031)}} & \shortstack{0.153$^{**}$\\{\scriptsize (0.045)}} & \shortstack{0.159$^{**}$\\{\scriptsize (0.048)}} & \shortstack{0.383\\{\scriptsize (0.048)}} & \shortstack{0.221\\{\scriptsize (0.083)}} & \shortstack{0.223\\{\scriptsize (0.083)}} \\
\texttt{astropt\_95m} & \shortstack{0.283$^{**}$\\{\scriptsize (0.050)}} & \shortstack{0.205$^{**}$\\{\scriptsize (0.059)}} & \shortstack{0.199$^{**}$\\{\scriptsize (0.093)}} & \shortstack{0.320$^{**}$\\{\scriptsize (0.024)}} & \shortstack{0.137$^{**}$\\{\scriptsize (0.036)}} & \shortstack{0.143$^{**}$\\{\scriptsize (0.039)}} & \shortstack{\textbf{0.352}\\{\scriptsize (0.040)}} & \shortstack{0.195\\{\scriptsize (0.069)}} & \shortstack{0.196\\{\scriptsize (0.069)}} \\
\texttt{astropt\_850m} & \shortstack{\textbf{0.246}$^{**}$\\{\scriptsize (0.041)}} & \shortstack{\textbf{0.171}$^{**}$\\{\scriptsize (0.047)}} & \shortstack{\textbf{0.169}$^{**}$\\{\scriptsize (0.077)}} & \shortstack{\textbf{0.313}$^{**}$\\{\scriptsize (0.016)}} & \shortstack{\textbf{0.111}$^{**}$\\{\scriptsize (0.029)}} & \shortstack{\textbf{0.116}$^{**}$\\{\scriptsize (0.031)}} & \shortstack{0.363\\{\scriptsize (0.031)}} & \shortstack{\textbf{0.159}\\{\scriptsize (0.054)}} & \shortstack{\textbf{0.160}\\{\scriptsize (0.054)}} \\
\addlinespace
\texttt{convnext\_nano} & \shortstack{0.611$^{**}$\\{\scriptsize (0.089)}} & \shortstack{0.472$^{**}$\\{\scriptsize (0.135)}} & \shortstack{0.413$^{**}$\\{\scriptsize (0.212)}} & \shortstack{0.506$^{**}$\\{\scriptsize (0.076)}} & \shortstack{0.348$^{**}$\\{\scriptsize (0.088)}} & \shortstack{0.384$^{**}$\\{\scriptsize (0.096)}} & \shortstack{0.587\\{\scriptsize (0.081)}} & \shortstack{0.467\\{\scriptsize (0.126)}} & \shortstack{0.470\\{\scriptsize (0.125)}} \\
\texttt{convnext\_tiny} & \shortstack{0.618$^{**}$\\{\scriptsize (0.086)}} & \shortstack{0.470$^{**}$\\{\scriptsize (0.134)}} & \shortstack{0.413$^{**}$\\{\scriptsize (0.222)}} & \shortstack{0.503$^{**}$\\{\scriptsize (0.072)}} & \shortstack{0.314$^{**}$\\{\scriptsize (0.091)}} & \shortstack{0.348$^{**}$\\{\scriptsize (0.100)}} & \shortstack{0.591\\{\scriptsize (0.070)}} & \shortstack{0.455\\{\scriptsize (0.137)}} & \shortstack{0.458\\{\scriptsize (0.137)}} \\
\texttt{convnext\_base} & \shortstack{0.588$^{**}$\\{\scriptsize (0.101)}} & \shortstack{0.474$^{**}$\\{\scriptsize (0.119)}} & \shortstack{0.415$^{**}$\\{\scriptsize (0.205)}} & \shortstack{0.514$^{**}$\\{\scriptsize (0.066)}} & \shortstack{0.330$^{**}$\\{\scriptsize (0.094)}} & \shortstack{0.359$^{**}$\\{\scriptsize (0.104)}} & \shortstack{0.566\\{\scriptsize (0.070)}} & \shortstack{0.434\\{\scriptsize (0.121)}} & \shortstack{0.436\\{\scriptsize (0.120)}} \\
\texttt{convnext\_large} & \shortstack{0.607$^{**}$\\{\scriptsize (0.088)}} & \shortstack{0.482$^{**}$\\{\scriptsize (0.128)}} & \shortstack{0.430$^{**}$\\{\scriptsize (0.212)}} & \shortstack{0.510$^{**}$\\{\scriptsize (0.070)}} & \shortstack{0.340$^{**}$\\{\scriptsize (0.094)}} & \shortstack{0.371$^{**}$\\{\scriptsize (0.104)}} & \shortstack{0.557\\{\scriptsize (0.069)}} & \shortstack{0.448\\{\scriptsize (0.109)}} & \shortstack{0.450\\{\scriptsize (0.109)}} \\
\addlinespace
\texttt{dino\_small} & \shortstack{0.560$^{**}$\\{\scriptsize (0.074)}} & \shortstack{0.447$^{**}$\\{\scriptsize (0.114)}} & \shortstack{0.388$^{**}$\\{\scriptsize (0.190)}} & \shortstack{0.447$^{**}$\\{\scriptsize (0.046)}} & \shortstack{0.263$^{**}$\\{\scriptsize (0.074)}} & \shortstack{0.280$^{**}$\\{\scriptsize (0.082)}} & \shortstack{0.495\\{\scriptsize (0.051)}} & \shortstack{0.339\\{\scriptsize (0.107)}} & \shortstack{0.341\\{\scriptsize (0.107)}} \\
\texttt{dino\_base} & \shortstack{0.612$^{**}$\\{\scriptsize (0.087)}} & \shortstack{0.496$^{**}$\\{\scriptsize (0.116)}} & \shortstack{0.442$^{**}$\\{\scriptsize (0.207)}} & \shortstack{0.439$^{**}$\\{\scriptsize (0.051)}} & \shortstack{0.264$^{**}$\\{\scriptsize (0.073)}} & \shortstack{0.281$^{**}$\\{\scriptsize (0.079)}} & \shortstack{0.525\\{\scriptsize (0.056)}} & \shortstack{0.336\\{\scriptsize (0.115)}} & \shortstack{0.339\\{\scriptsize (0.115)}} \\
\texttt{dino\_large} & \shortstack{0.610$^{**}$\\{\scriptsize (0.089)}} & \shortstack{0.468$^{**}$\\{\scriptsize (0.129)}} & \shortstack{0.410$^{**}$\\{\scriptsize (0.205)}} & \shortstack{0.450\\{\scriptsize (0.043)}} & \shortstack{0.241$^{**}$\\{\scriptsize (0.068)}} & \shortstack{0.258$^{**}$\\{\scriptsize (0.076)}} & \shortstack{0.514\\{\scriptsize (0.045)}} & \shortstack{0.319\\{\scriptsize (0.104)}} & \shortstack{0.321\\{\scriptsize (0.104)}} \\
\texttt{dino\_giant} & \shortstack{0.608\\{\scriptsize (0.090)}} & \shortstack{0.459$^{**}$\\{\scriptsize (0.142)}} & \shortstack{0.404$^{**}$\\{\scriptsize (0.209)}} & \shortstack{0.403$^{**}$\\{\scriptsize (0.041)}} & \shortstack{0.195$^{**}$\\{\scriptsize (0.062)}} & \shortstack{0.210$^{**}$\\{\scriptsize (0.069)}} & \shortstack{0.536\\{\scriptsize (0.044)}} & \shortstack{0.263\\{\scriptsize (0.100)}} & \shortstack{0.265\\{\scriptsize (0.100)}} \\
\addlinespace
\texttt{ijepa\_giant} & \shortstack{0.506$^{**}$\\{\scriptsize (0.089)}} & \shortstack{0.371$^{**}$\\{\scriptsize (0.128)}} & \shortstack{0.342$^{**}$\\{\scriptsize (0.199)}} & \shortstack{0.450$^{**}$\\{\scriptsize (0.039)}} & \shortstack{0.188$^{**}$\\{\scriptsize (0.093)}} & \shortstack{0.200$^{**}$\\{\scriptsize (0.095)}} & \shortstack{0.436\\{\scriptsize (0.056)}} & \shortstack{0.288\\{\scriptsize (0.119)}} & \shortstack{0.290\\{\scriptsize (0.120)}} \\
\texttt{ijepa\_huge} & \shortstack{0.405$^{**}$\\{\scriptsize (0.069)}} & \shortstack{0.277$^{**}$\\{\scriptsize (0.095)}} & \shortstack{0.257$^{**}$\\{\scriptsize (0.153)}} & \shortstack{0.449$^{**}$\\{\scriptsize (0.037)}} & \shortstack{0.191$^{**}$\\{\scriptsize (0.092)}} & \shortstack{0.204$^{**}$\\{\scriptsize (0.097)}} & \shortstack{0.450\\{\scriptsize (0.054)}} & \shortstack{0.274\\{\scriptsize (0.118)}} & \shortstack{0.276\\{\scriptsize (0.118)}} \\
\addlinespace
\texttt{vit\_base} & \shortstack{0.725$^{**}$\\{\scriptsize (0.076)}} & \shortstack{0.597$^{**}$\\{\scriptsize (0.127)}} & \shortstack{0.518$^{**}$\\{\scriptsize (0.241)}} & \shortstack{0.549\\{\scriptsize (0.054)}} & \shortstack{0.352$^{**}$\\{\scriptsize (0.085)}} & \shortstack{0.377$^{**}$\\{\scriptsize (0.093)}} & \shortstack{0.573$^{**}$\\{\scriptsize (0.067)}} & \shortstack{0.446\\{\scriptsize (0.115)}} & \shortstack{0.448\\{\scriptsize (0.115)}} \\
\texttt{vit\_large} & \shortstack{0.496$^{**}$\\{\scriptsize (0.059)}} & \shortstack{0.430$^{**}$\\{\scriptsize (0.076)}} & \shortstack{0.380$^{**}$\\{\scriptsize (0.165)}} & \shortstack{0.369$^{**}$\\{\scriptsize (0.032)}} & \shortstack{0.266$^{**}$\\{\scriptsize (0.049)}} & \shortstack{0.280$^{**}$\\{\scriptsize (0.052)}} & \shortstack{0.400$^{**}$\\{\scriptsize (0.036)}} & \shortstack{0.331\\{\scriptsize (0.069)}} & \shortstack{0.333\\{\scriptsize (0.069)}} \\
\texttt{vit\_huge} & \shortstack{0.403$^{**}$\\{\scriptsize (0.049)}} & \shortstack{0.287$^{**}$\\{\scriptsize (0.086)}} & \shortstack{0.259$^{**}$\\{\scriptsize (0.132)}} & \shortstack{0.368$^{**}$\\{\scriptsize (0.028)}} & \shortstack{0.206$^{**}$\\{\scriptsize (0.048)}} & \shortstack{0.220$^{**}$\\{\scriptsize (0.052)}} & \shortstack{0.394\\{\scriptsize (0.036)}} & \shortstack{0.292\\{\scriptsize (0.083)}} & \shortstack{0.294\\{\scriptsize (0.084)}} \\
\bottomrule
\end{tabular}
\end{table}

\begin{table}[htbp]
\centering
\caption{Median normalized cross-survey translation residual for three translator classes.
HSC$\rightarrow$JWST and Legacy$\rightarrow$HSC use matched source and target backbone families.
For SDSS$\rightarrow$HSC, the source representation is fixed to \texttt{specformer\_base\_sdss}.
Residuals should therefore be compared within, rather than across, survey pairs.
All entries are L2 residual norms. Lower is better; the best value in each column is bold.}
\label{tab:residuals-median}
\small
\begin{tabular}{lccccccccc}
\toprule
& \multicolumn{3}{c}{HSC $\rightarrow$ JWST ($n$=1.5K)}
& \multicolumn{3}{c}{Legacy $\rightarrow$ HSC ($n$=100K)}
& \multicolumn{3}{c}{SDSS $\rightarrow$ HSC ($n$=2K)} \\
\cmidrule(lr){2-4} \cmidrule(lr){5-7} \cmidrule(lr){8-10}
Backbone & SVR & Ridge & $k$-NN & SVR & Ridge & $k$-NN & SVR & Ridge & $k$-NN \\
\midrule
\texttt{astropt\_15m}  & 0.267 & 0.229 & 0.221 & 0.197 & 0.135 & 0.146 & 0.255 & 0.206 & 0.198 \\
\texttt{astropt\_95m}  & 0.240 & 0.197 & 0.195 & \textbf{0.183} & 0.121 & 0.128 & 0.231 & 0.184 & 0.177 \\
\texttt{astropt\_850m} & \textbf{0.196} & \textbf{0.168} & \textbf{0.167} & 0.202 & \textbf{0.098} & \textbf{0.105} & \textbf{0.213} & \textbf{0.148} & \textbf{0.141} \\
\midrule
\texttt{convnext\_nano}  & 0.509 & 0.466 & 0.405 & 0.442 & 0.316 & 0.316 & 0.453 & 0.446 & 0.444 \\
\texttt{convnext\_tiny}  & 0.490 & 0.462 & 0.407 & 0.375 & 0.280 & 0.314 & 0.428 & 0.421 & 0.416 \\
\texttt{convnext\_base}  & 0.517 & 0.465 & 0.424 & 0.384 & 0.292 & 0.284 & 0.417 & 0.406 & 0.410 \\
\texttt{convnext\_large} & 0.532 & 0.470 & 0.443 & 0.396 & 0.301 & 0.266 & 0.440 & 0.431 & 0.432 \\
\midrule
\texttt{dino\_small} & 0.492 & 0.437 & 0.400 & 0.427 & 0.236 & 0.242 & 0.353 & 0.316 & 0.309 \\
\texttt{dino\_base}  & 0.550 & 0.497 & 0.468 & 0.325 & 0.238 & 0.239 & 0.329 & 0.312 & 0.308 \\
\texttt{dino\_large} & 0.507 & 0.462 & 0.419 & 0.286 & 0.217 & 0.205 & 0.315 & 0.302 & 0.292 \\
\texttt{dino\_giant} & 0.493 & 0.452 & 0.399 & 0.229 & 0.173 & 0.159 & 0.248 & 0.245 & 0.235 \\
\midrule
\texttt{ijepa\_giant} & 0.395 & 0.363 & 0.318 & 0.317 & 0.145 & 0.126 & 0.275 & 0.263 & 0.244 \\
\texttt{ijepa\_huge}  & 0.292 & 0.266 & 0.232 & 0.285 & 0.150 & 0.150 & 0.266 & 0.244 & 0.232 \\
\midrule
\texttt{vit\_base}  & 0.644 & 0.583 & 0.552 & 0.437 & 0.326 & 0.326 & 0.431 & 0.426 & 0.428 \\
\texttt{vit\_large} & 0.452 & 0.427 & 0.408 & 0.291 & 0.248 & 0.225 & 0.321 & 0.322 & 0.314 \\
\texttt{vit\_huge}  & 0.307 & 0.284 & 0.255 & 0.256 & 0.187 & 0.181 & 0.281 & 0.279 & 0.271 \\
\bottomrule
\end{tabular}
\end{table}

\begin{table}[htbp]
\centering
\caption{90th-percentile normalized cross-survey translation residual for three translator classes.
HSC$\rightarrow$JWST and Legacy$\rightarrow$HSC use matched source and target backbone families.
For SDSS$\rightarrow$HSC, the source representation is fixed to \texttt{specformer\_base\_sdss}.
Residuals should therefore be compared within, rather than across, survey pairs.
All entries are L2 residual norms. Lower is better; the best value in each column is bold.}
\label{tab:residuals-p90}
\small
\begin{tabular}{lccccccccc}
\toprule
& \multicolumn{3}{c}{HSC $\rightarrow$ JWST ($n$=1.5K)}
& \multicolumn{3}{c}{Legacy $\rightarrow$ HSC ($n$=100K)}
& \multicolumn{3}{c}{SDSS $\rightarrow$ HSC ($n$=2K)} \\
\cmidrule(lr){2-4} \cmidrule(lr){5-7} \cmidrule(lr){8-10}
Backbone & SVR & Ridge & $k$-NN & SVR & Ridge & $k$-NN & SVR & Ridge & $k$-NN \\
\midrule
\texttt{astropt\_15m}  & 0.386 & 0.362 & 0.334 & 0.272 & 0.211 & 0.210 & 0.353 & 0.352 & 0.341 \\
\texttt{astropt\_95m}  & 0.330 & 0.299 & 0.285 & \textbf{0.243} & 0.180 & 0.182 & 0.304 & 0.300 & 0.287 \\
\texttt{astropt\_850m} & \textbf{0.282} & \textbf{0.247} & \textbf{0.234} & 0.243 & \textbf{0.143} & \textbf{0.154} & \textbf{0.285} & \textbf{0.248} & \textbf{0.232} \\
\midrule
\texttt{convnext\_nano}  & 0.723 & 0.637 & 0.671 & 0.577 & 0.432 & 0.429 & 0.666 & 0.641 & 0.654 \\
\texttt{convnext\_tiny}  & 0.743 & 0.658 & 0.681 & 0.550 & 0.404 & 0.446 & 0.675 & 0.660 & 0.654 \\
\texttt{convnext\_base}  & 0.691 & 0.625 & 0.607 & 0.544 & 0.421 & 0.409 & 0.610 & 0.586 & 0.597 \\
\texttt{convnext\_large} & 0.705 & 0.631 & 0.643 & 0.561 & 0.435 & 0.397 & 0.612 & 0.595 & 0.595 \\
\midrule
\texttt{dino\_small} & 0.657 & 0.591 & 0.590 & 0.502 & 0.341 & 0.346 & 0.513 & 0.506 & 0.497 \\
\texttt{dino\_base}  & 0.701 & 0.645 & 0.653 & 0.446 & 0.341 & 0.346 & 0.518 & 0.503 & 0.487 \\
\texttt{dino\_large} & 0.707 & 0.636 & 0.641 & 0.410 & 0.317 & 0.293 & 0.488 & 0.476 & 0.455 \\
\texttt{dino\_giant} & 0.711 & 0.651 & 0.635 & 0.346 & 0.264 & 0.245 & 0.403 & 0.403 & 0.384 \\
\midrule
\texttt{ijepa\_giant} & 0.610 & 0.585 & 0.588 & 0.460 & 0.307 & 0.204 & 0.466 & 0.463 & 0.446 \\
\texttt{ijepa\_huge}  & 0.443 & 0.419 & 0.399 & 0.448 & 0.310 & 0.269 & 0.454 & 0.442 & 0.413 \\
\midrule
\texttt{vit\_base}  & 0.795 & 0.718 & 0.751 & 0.569 & 0.445 & 0.471 & 0.627 & 0.606 & 0.607 \\
\texttt{vit\_large} & 0.566 & 0.519 & 0.528 & 0.373 & 0.314 & 0.303 & 0.426 & 0.428 & 0.415 \\
\texttt{vit\_huge}  & 0.428 & 0.399 & 0.382 & 0.339 & 0.258 & 0.248 & 0.425 & 0.430 & 0.404 \\
\bottomrule
\end{tabular}
\end{table}

\subsection{Global conformal calibration}
\label{app:global_calibration}

\begin{figure*}[htbp]
    \centering
    \includegraphics[width=1.0\linewidth]
    {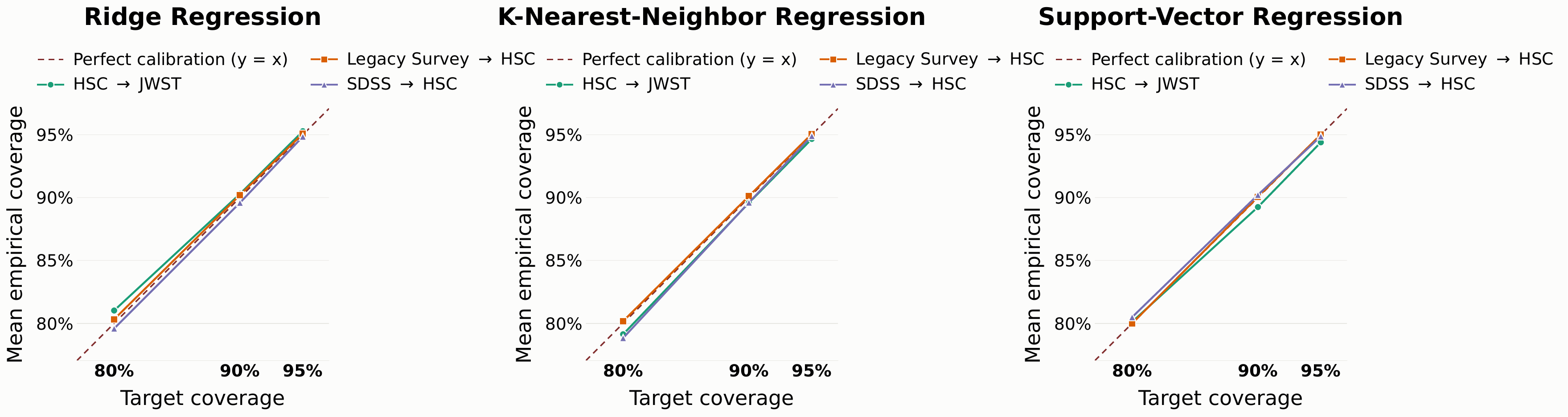}
    \caption{Empirical test coverage of the global conformal regions at target coverages of $80\%$, $90\%$, and $95\%$. Coverage remains close to nominal for all three survey pairs.}
    \label{fig:global_cp_calibration}
\end{figure*}

\begin{figure*}[htbp]
    \centering
    \includegraphics[width=1.0\linewidth]
    {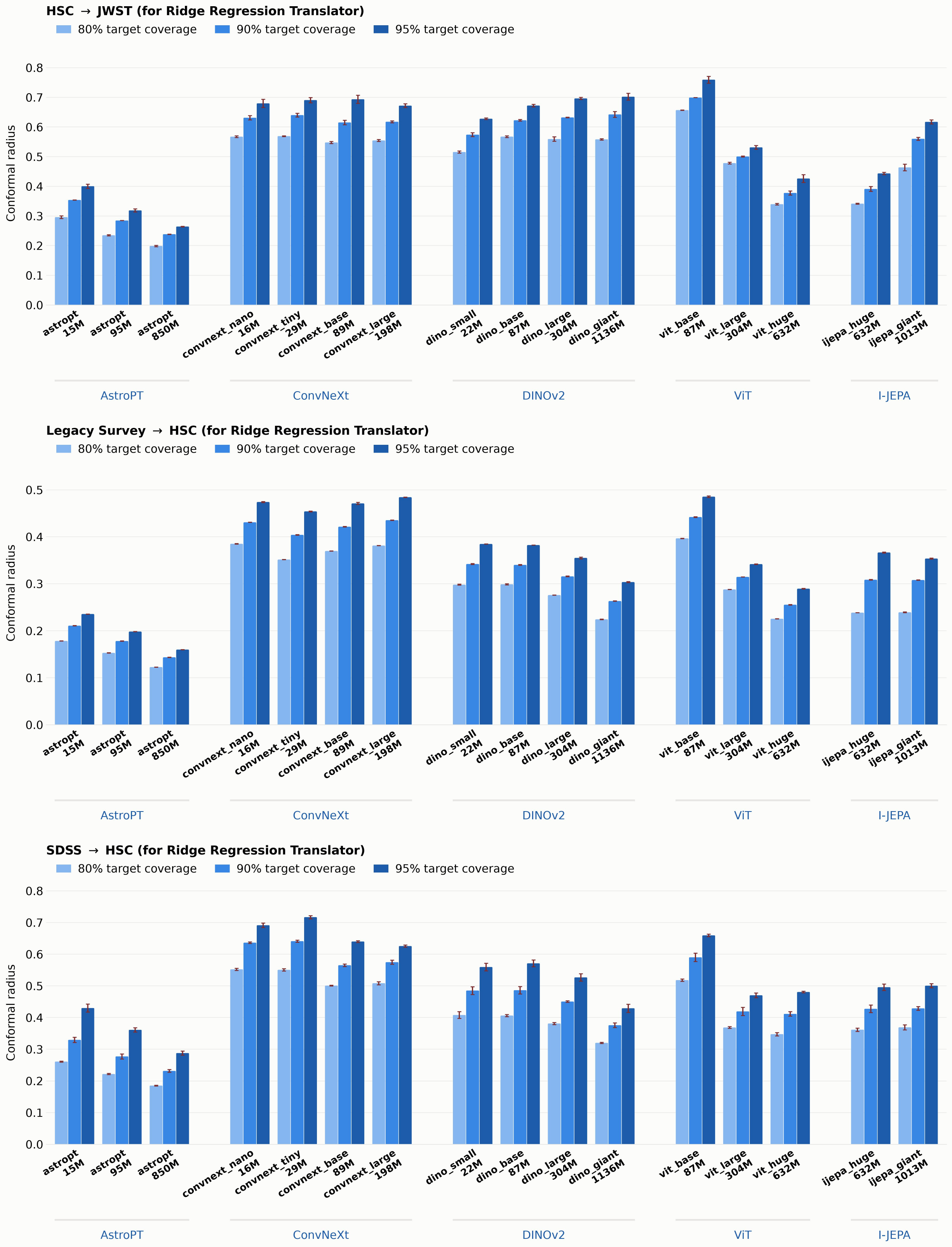}
    \caption{
    Conformal radius required at $80\%$, $90\%$, and $95\%$ target coverage
    for all three survey-pairs using \textbf{ridge regression} translator. Error bars show $\pm1$ standard
    deviation across three calibration/test reshuffles. Smaller regions at
    the same coverage indicate more precise cross-survey transfer.
    }
    \label{fig:global_cp_radius_ridge}
\end{figure*}

\begin{figure*}[htbp]
    \centering
    \includegraphics[width=1.0\linewidth]
    {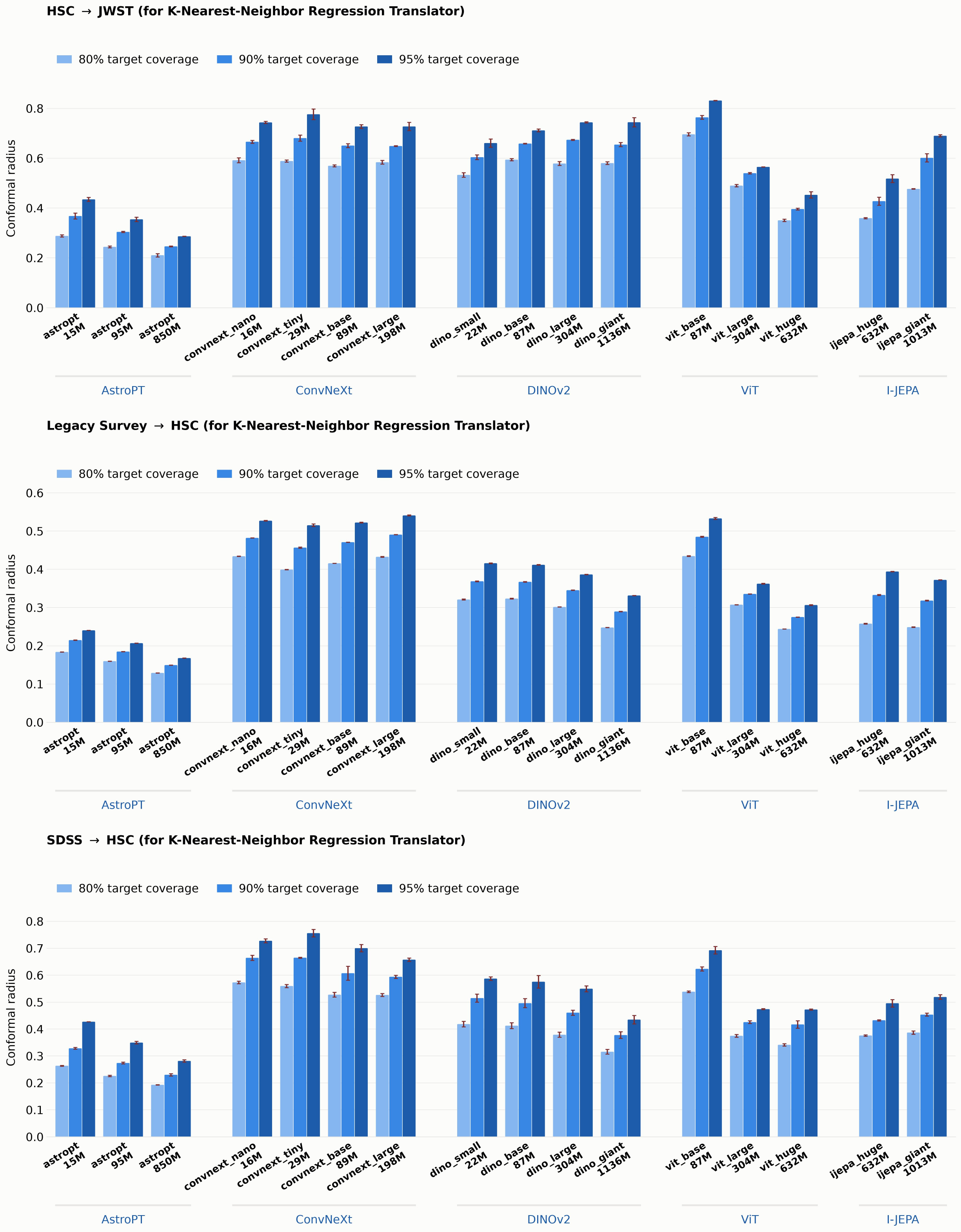}
    \caption{
    Conformal radius required at $80\%$, $90\%$, and $95\%$ target coverage
    for all three survey-pairs using \textbf{$k$-nearest-neighbor} translator. Error bars show $\pm1$ standard
    deviation across three calibration/test reshuffles. Smaller regions at
    the same coverage indicate more precise cross-survey transfer.
    }
    \label{fig:global_cp_radius_knn}
\end{figure*}

\begin{figure*}[htbp]
    \centering
    \includegraphics[width=1.0\linewidth]
    {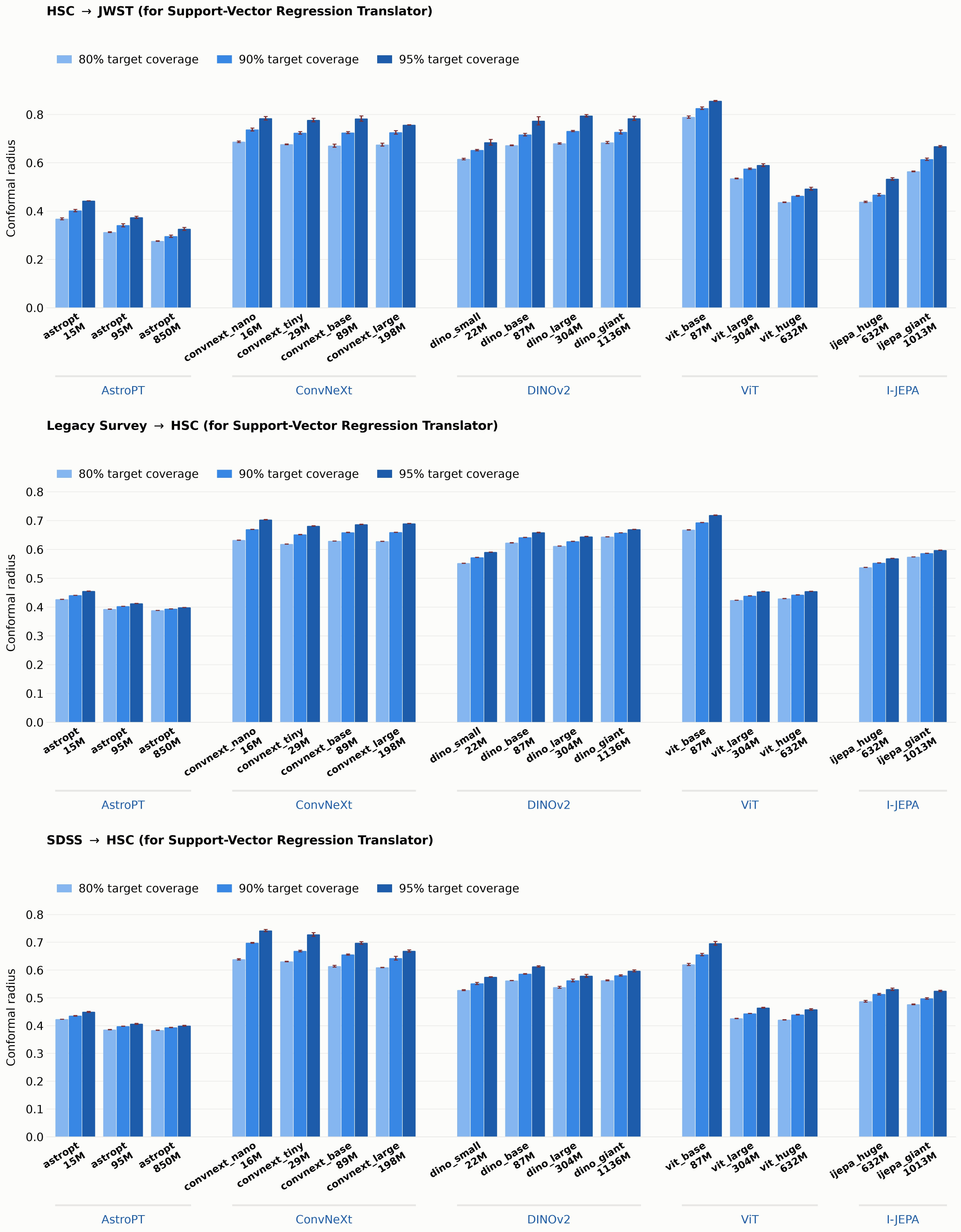}
    \caption{
    Conformal radius required at $80\%$, $90\%$, and $95\%$ target coverage
    for all three survey-pairs using \textbf{radial basis function (RBF) support-vector regression} translator. Error bars show $\pm1$ standard
    deviation across three calibration/test reshuffles. Smaller regions at
    the same coverage indicate more precise cross-survey transfer.
    }
    \label{fig:global_cp_radius_svr}
\end{figure*}

The global conformal regions are well calibrated across all three survey
pairs. Figure~\ref{fig:global_cp_calibration} shows the empirical coverage
at target levels of $80\%$, $90\%$, and $95\%$ using translators: ridge regression, $k$-nearest neighbor regression and radial basis function (RBF) support-vector regression. The measured coverages
remain close to the requested levels for HSC$\rightarrow$JWST,
Legacy Survey$\rightarrow$HSC, and SDSS$\rightarrow$HSC. This check
separates calibration from efficiency: the models attain similar marginal
coverage, while the conformal radii reported in Figures~\ref{fig:global_cp_radius_ridge}--~\ref{fig:global_cp_radius_svr} can differ
substantially.


\subsection{Adaptive coverage across predicted-difficulty groups}
\label{app:adaptive_quartiles}

\begin{figure}[htbp]
    \centering
    \includegraphics[width=\linewidth]{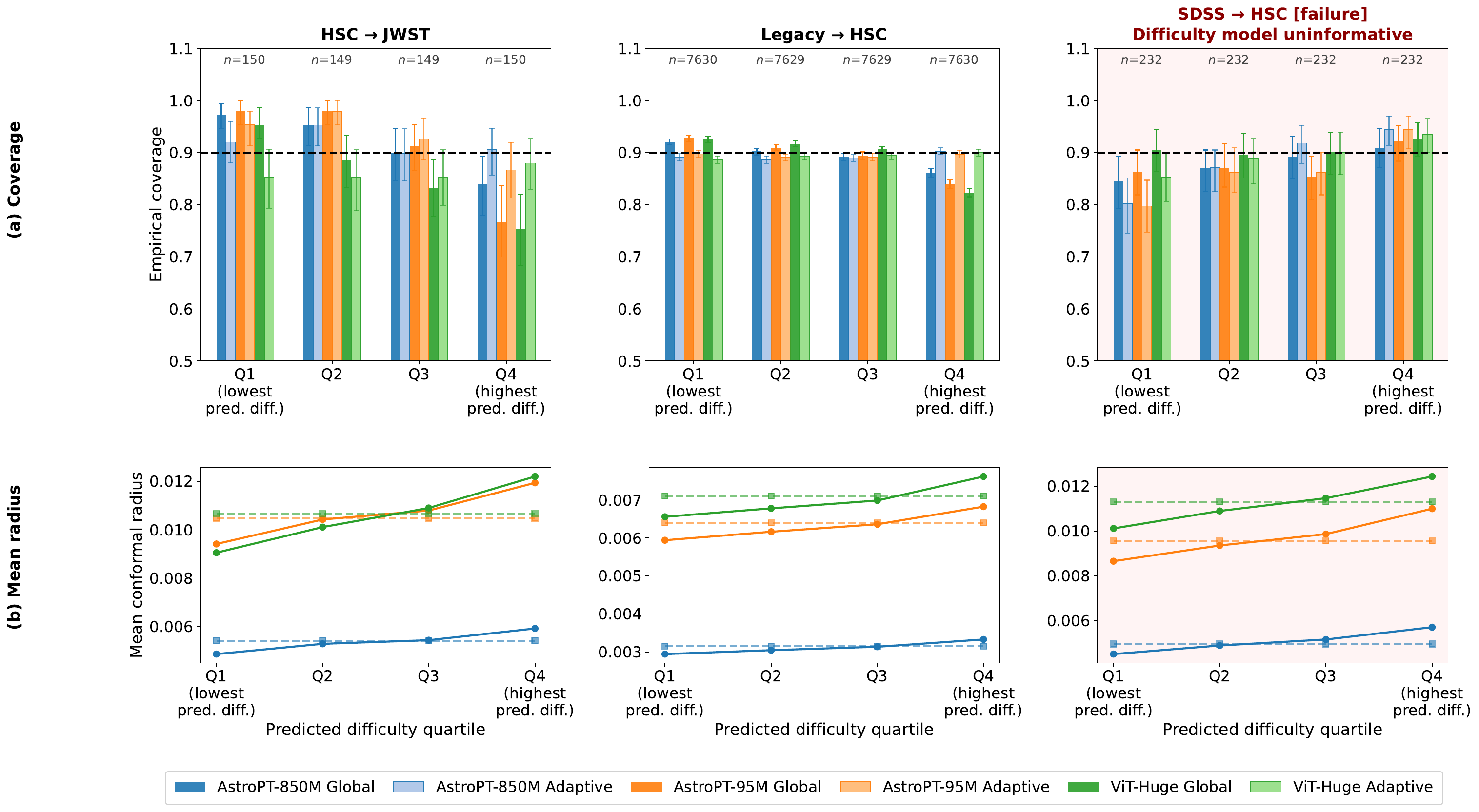}
    \caption{Coverage by predicted-difficulty quartile at $90\%$ target coverage for AstroPT-850M, AstroPT-95M, and ViT-Huge. \textbf{(a)} Empirical coverage of global (dark) and adaptive (light) conformal regions; error bars are $95\%$ bootstrap confidence intervals and $n$ denotes the number of test objects per quartile. \textbf{(b)} Mean conformal radius per quartile; solid lines show the adaptive radius and dashed lines show the constant global radius. Adaptive calibration increases region sizes for objects with higher predicted difficulty, improving Q4 coverage when the difficulty signal is informative (HSC$\rightarrow$JWST, Legacy$\rightarrow$HSC). SDSS$\rightarrow$HSC provides a negative-control case: because predicted difficulty is uninformative there, adaptive radii still grow across quartiles but coverage does not consistently improve.}
    \label{fig:quartile_coverage}
\end{figure}

Global conformal prediction is calibrated marginally, but this does not guarantee equal coverage for objects of different predicted difficulty. Figure~\ref{fig:quartile_coverage} compares global and adaptive coverage after grouping test objects into quartiles using the source-side difficulty estimate. When predicted difficulty is informative, adaptive calibration moves coverage toward the nominal $90\%$ level for the hardest objects. The effect is strongest for Legacy Survey$\rightarrow$HSC and for selected HSC$\rightarrow$JWST
backbones, while SDSS$\rightarrow$HSC shows little consistent benefit.

\subsection{Failure-set overlap across survey pairs and translators}
\label{app:failure_overlap}

\begin{figure*}[htbp]
\centering

\begin{minipage}[htbp]{0.47\linewidth}
    \centering
    \includegraphics[width=\linewidth]
    {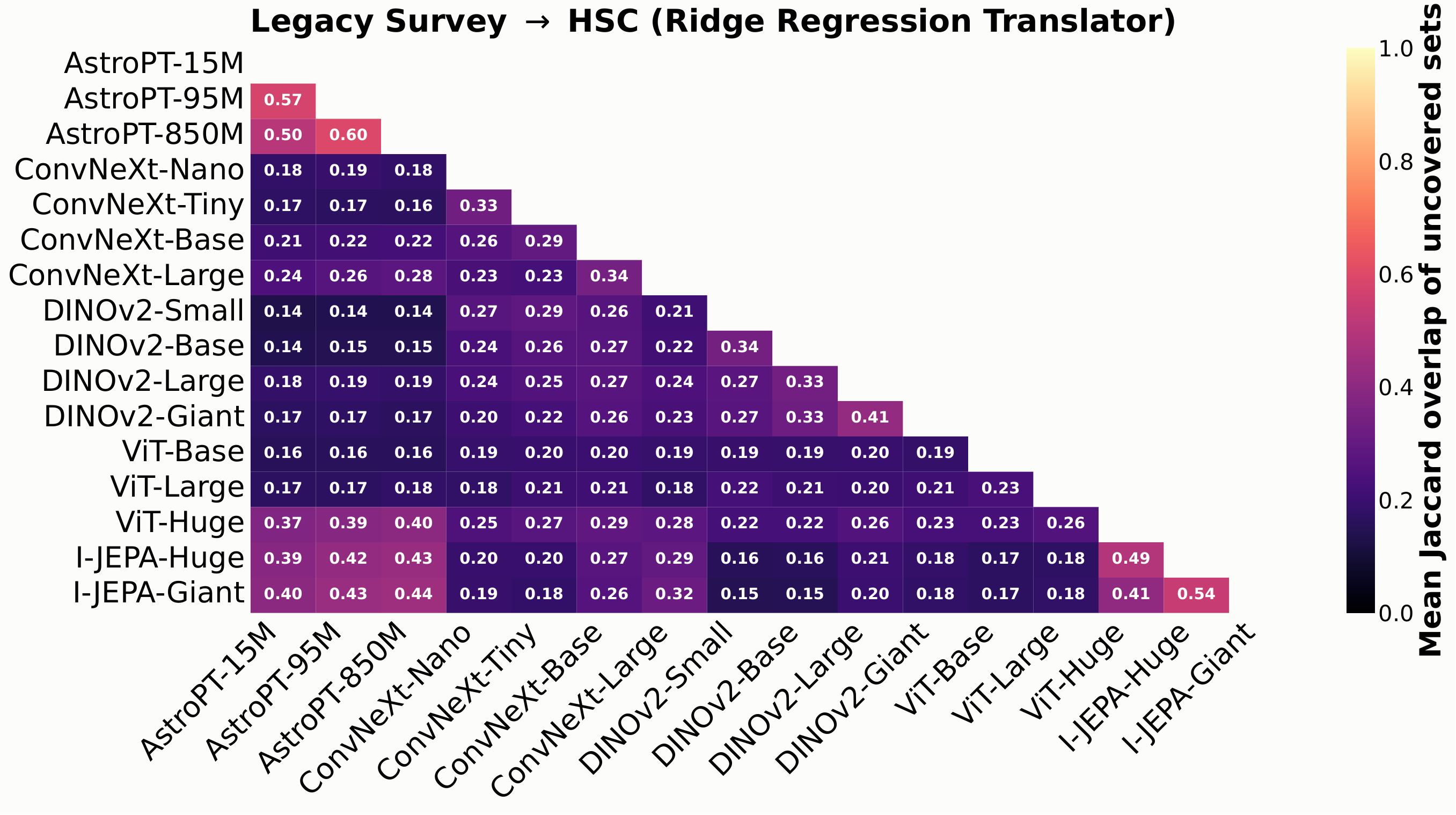}

    \vspace{-1mm}
    {\small (a) Legacy Survey $\rightarrow$ HSC}
\end{minipage}
\hfill
\begin{minipage}[htbp]{0.47\linewidth}
    \centering
    \includegraphics[width=\linewidth]
    {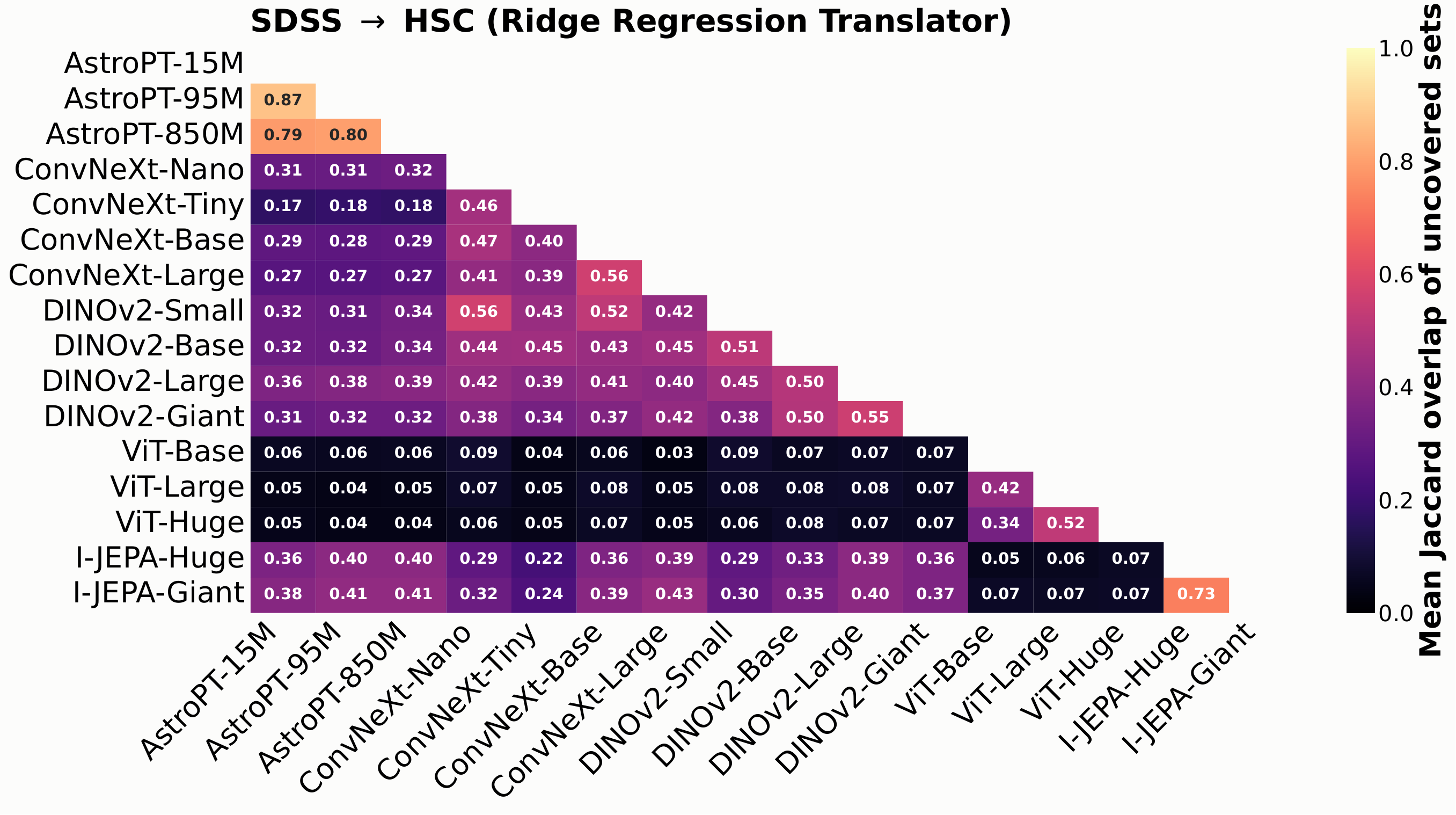}

    \vspace{-1mm}
    {\small (b) SDSS $\rightarrow$ HSC}
\end{minipage}

\caption{
Pairwise Jaccard overlap of uncovered-object sets at $90\%$ target
coverage for the two additional survey pairs. Higher values indicate
that two representations fail on a larger fraction of the same objects.
}
\label{fig:ridge_additional_fsilure_overlap}
\end{figure*}

\begin{figure*}[htbp]
    \centering
    \includegraphics[width=1.0\linewidth]
    {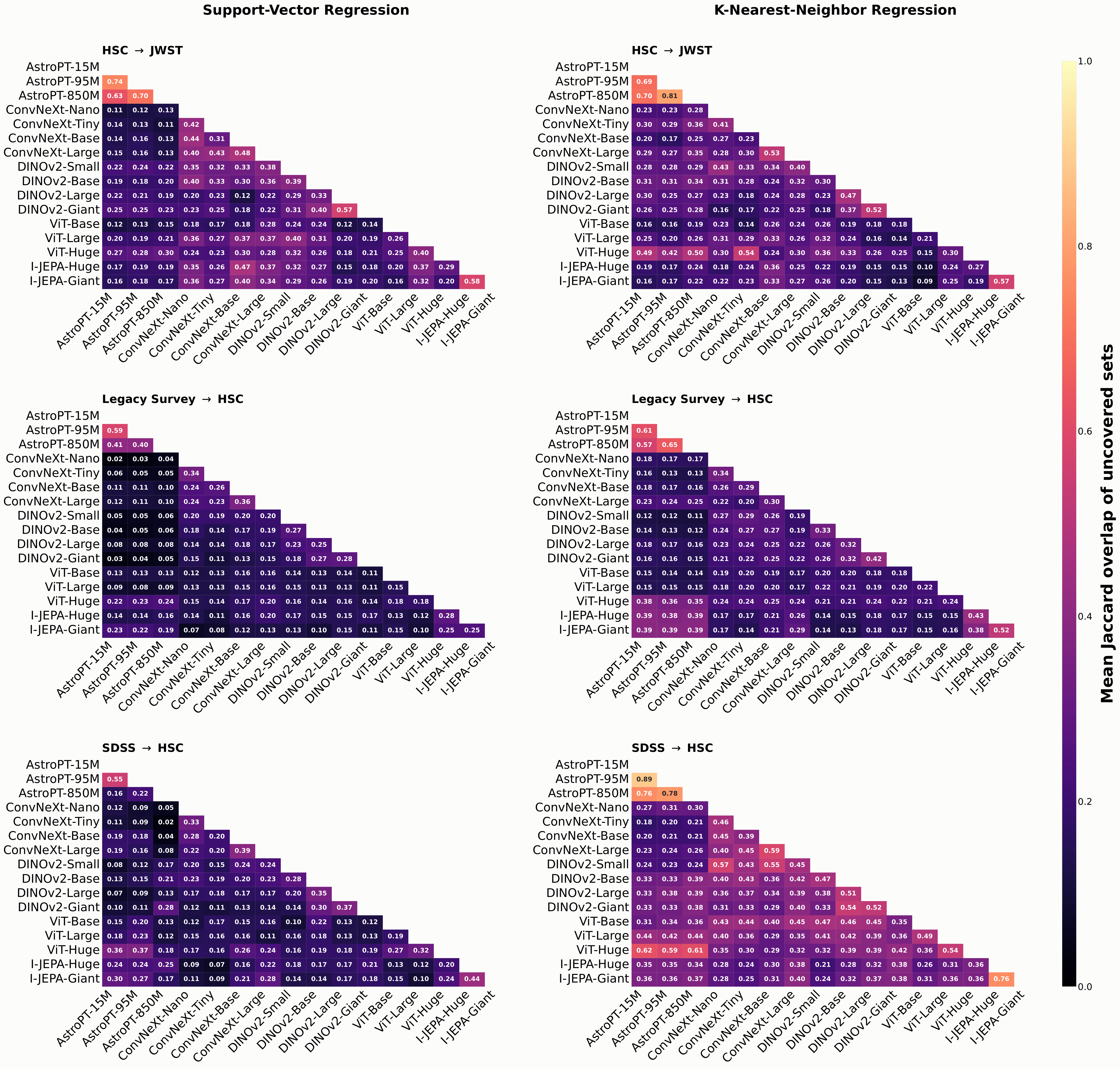}
\caption{
Pairwise Jaccard indices between uncovered-object sets at $90\%$
target
coverage for HSC $\rightarrow$JWST (top), Legacy Survey$\rightarrow$HSC
(middle), and SDSS$\rightarrow$HSC (bottom), using support vector regression
(left) and $k$-nearest-neighbor regression (right). Each cell reports the mean
across three calibration/test reshuffles.
Higher values indicate greater similarity between the corresponding failure
sets.
}
    \label{fig:other_additional_fsilure_overlap}
\end{figure*}

The main text reports the HSC$\rightarrow$JWST failure-set overlap at $90\%$ target coverage for ridge regression translator. Figure~\ref{fig:ridge_additional_fsilure_overlap} shows the corresponding matrices for Legacy Survey$\rightarrow$HSC and SDSS$\rightarrow$HSC using ridge regression translator. The overlap structure changes with the survey pair, consistent with the observation that cross-survey failures depend on both the representation and the source--target setting. We also report the overlap matrices for the other translators used in Figure~\ref{fig:other_additional_fsilure_overlap}.

\section{Adaptive conformal failure categories}
\label{app:failure_types}

We examine the Legacy Survey$\rightarrow$HSC AstroPT-850M setting in greater detail because it provides a clear difficulty signal and has the largest matched sample. At $90\%$ target coverage, an \emph{adaptive rescue} is missed by global conformal prediction but covered after adaptation; a \emph{silent failure} has low predicted difficulty despite a large residual; a \emph{false alarm} is predicted to be difficult despite a small residual; and an \emph{adaptive regression} is globally covered but becomes uncovered after adaptation. Matched examples compare each flagged object with a control while preserving pair identity across Legacy Survey and HSC images.




\begin{figure}[htbp]
\centering

\threepairgrid{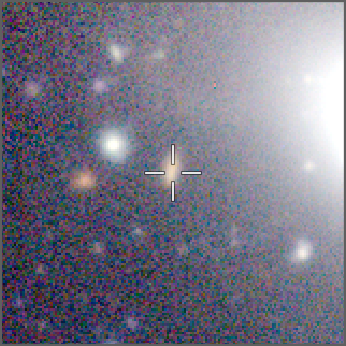}

\vspace{-1mm}

\begingroup
\centering
\footnotesize
\setlength{\tabcolsep}{0.9em}
\renewcommand{\arraystretch}{1.08}
\begin{tabular}{@{}lccc@{}}
\toprule
 & \textbf{Pair 1} & \textbf{Pair 2} & \textbf{Pair 3} \\
\midrule
$\Delta r$ & +0.0015 & +0.0015 & +0.0025 \\
$\Delta R^{\mathrm g}$ & +0.0000 & +0.0000 & +0.0000 \\
$\Delta \mathrm{PSF}\;[^\prime\prime]$ & +0.07 & +0.07 & +0.15 \\
\bottomrule
\end{tabular}
\endgroup

\caption{Representative adaptive rescues and matched easy-stable controls.
Each row corresponds to one matched pair and shows both Legacy Survey
and HSC imaging. Crosshairs mark the target object. Flagged and control
images within each survey use the same display normalization.
}

\label{fig:adaptive_rescue_examples}
\end{figure}


\begin{figure}[htbp]
\centering

\threepairgrid{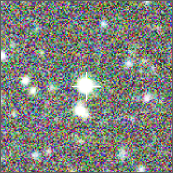}

\vspace{-1mm}

\begingroup
\centering
\footnotesize
\setlength{\tabcolsep}{0.9em}
\renewcommand{\arraystretch}{1.08}
\begin{tabular}{@{}lccc@{}}
\toprule
 & \textbf{Pair 1} & \textbf{Pair 2} & \textbf{Pair 3} \\
\midrule
$\Delta r$ & +0.0019 & +0.0017 & +0.0022 \\
$\Delta R^{\mathrm g}$ & +0.0000 & +0.0000 & +0.0000 \\
$\Delta \mathrm{PSF}\;[^\prime\prime]$ & +0.04 & +0.08 & +0.32 \\
\bottomrule
\end{tabular}
\endgroup

\caption{Representative silent failures and matched easy-stable controls. These objects have relatively low predicted difficulty despite large cross-survey residuals. Each row preserves the same matched pair across Legacy Survey and HSC imaging.
}
\label{fig:silent_failure_examples}
\end{figure}


\begin{figure}[htbp]
\centering

\threepairgrid{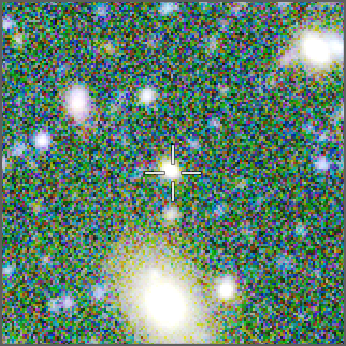}

\vspace{-1mm}

\begingroup
\centering
\footnotesize
\setlength{\tabcolsep}{0.9em}
\renewcommand{\arraystretch}{1.08}
\begin{tabular}{@{}lccc@{}}
\toprule
 & \textbf{Pair 1} & \textbf{Pair 2} & \textbf{Pair 3} \\
\midrule
$\Delta r$ & +0.0001 & -0.0001 & +0.0000 \\
$\Delta R^{\mathrm g}$ & +0.0000 & +0.0000 & +0.0000 \\
$\Delta \mathrm{PSF}\;[^\prime\prime]$ & +0.13 & +0.11 & +0.02 \\
\bottomrule
\end{tabular}
\endgroup

\caption{
Representative false alarms and their matched comparison objects.
The difficulty model assigns these objects high predicted difficulty,
although their observed cross-survey residuals are comparatively small.
}
\label{fig:false_alarm_examples}
\end{figure}


\begin{figure}[htbp]
\centering

\threepairgrid{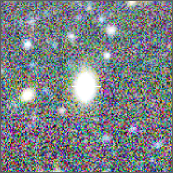}

\vspace{-1mm}

\begingroup
\centering
\footnotesize
\setlength{\tabcolsep}{0.9em}
\renewcommand{\arraystretch}{1.08}
\begin{tabular}{@{}lccc@{}}
\toprule
 & \textbf{Pair 1} & \textbf{Pair 2} & \textbf{Pair 3} \\
\midrule
$\Delta r$ & +0.0013 & +0.0012 & +0.0013 \\
$\Delta R^{\mathrm g}$ & +0.0000 & +0.0000 & +0.0000 \\
$\Delta \mathrm{PSF}\;[^\prime\prime]$ & +0.03 & +0.00 & +0.05 \\
\bottomrule
\end{tabular}
\endgroup

\caption{
Representative adaptive regressions and matched comparison objects.
These objects are covered by the global conformal region but become
uncovered after the radius is adapted to predicted difficulty.
}
\label{fig:adaptive_regression_examples}
\end{figure}

\FloatBarrier

\section{Matched observable comparisons}
\label{app:matched_stats}

We also tested whether the adaptive categories can be separated using standard observable quantities. The matched comparisons use $i$-band magnitude,
apparent size, and nearest-neighbor distance, none of which enters the category
definitions. Residual differences are included for reference, although the
residual contributes directly to the conformal outcomes.

For each quantity, we report the median paired difference and a $95\%$ bootstrap confidence interval \citep{efron1993introduction}. Statistical significance is assessed with two-sided paired sign-flip permutation tests, followed by Holm correction over the 12 observable comparisons. The results can be found in Table~\ref{tab:adaptive_effect}.

\begin{table*}[htbp]
\centering
\caption{
Matched comparison of adaptive-conformal object categories for
Legacy Survey$\rightarrow$HSC with AstroPT-850M at $90\%$ target coverage.
$\Delta$ is the median flagged-minus-comparison difference and intervals are
$95\%$ bootstrap confidence intervals from 2000 paired resamples.
$p_{\rm Holm}$ denotes the Holm-corrected two-sided paired permutation
$p$-value. Residual differences are reported descriptively because residual
information contributes to the category definitions. Bold entries indicate
$p_{\rm Holm}<0.05$.
}
\label{tab:adaptive_effect}

\scriptsize
\setlength{\tabcolsep}{4pt}
\renewcommand{\arraystretch}{1.08}

\begin{tabular}{llrrr}
\toprule
Comparison & Quantity & $\Delta$ median & 95\% CI & $p_{\rm Holm}$ \\
\midrule

\multirow{4}{*}{Rescue vs.\ stable}
 & $i$-mag
 & $+1.20$
 & $[+0.03,+1.96]$
 & 0.336 \\

 & SHAPE\_R [$''$]
 & $-0.29$
 & $[-0.82,+0.55]$
 & 0.900 \\

 & \textbf{NN distance}
 & \textbf{+0.0183}
 & \textbf{[+0.0162,+0.0203]}
 & \textbf{$<0.001$} \\

 & residual $r$
 & $+0.0015$
 & $[+0.00147,+0.00152]$
 & -- \\
\midrule

\multirow{4}{*}{Silent vs.\ stable}
 & $i$-mag
 & $+0.04$
 & $[-0.71,+1.50]$
 & 0.920 \\

 & SHAPE\_R [$''$]
 & $-0.13$
 & $[-0.59,+0.16]$
 & 0.920 \\

 & NN distance
 & $+0.0006$
 & $[-0.0001,+0.0011]$
 & 0.920 \\

 & residual $r$
 & $+0.0018$
 & $[+0.00170,+0.00179]$
 & -- \\
\midrule

\multirow{4}{*}{False alarm vs.\ hard}
 & $i$-mag
 & $+0.52$
 & $[-0.51,+0.94]$
 & 0.720 \\

 & SHAPE\_R [$''$]
 & $-0.34$
 & $[-1.09,+0.28]$
 & 0.920 \\

 & \textbf{NN distance}
 & \textbf{-0.0089}
 & \textbf{[-0.0092,-0.0045]}
 & \textbf{$<0.001$} \\

 & residual $r$
 & $-0.0008$
 & $[-0.00082,-0.00074]$
 & -- \\
\bottomrule
\end{tabular}
\end{table*}

\section{PSF-matching check}
\label{app:psf_match}

HSC has finer angular resolution than the Legacy Survey, so some of the differences visible in the matched images may arise from the survey PSFs rather than from the underlying source. We therefore convolve selected HSC cutouts to an approximate Legacy Survey PSF FWHM of $1.2''$ and compare the original and degraded HSC images with the corresponding Legacy cutout.

\begin{figure}[t]
\centering

\includegraphics[width=\linewidth]
{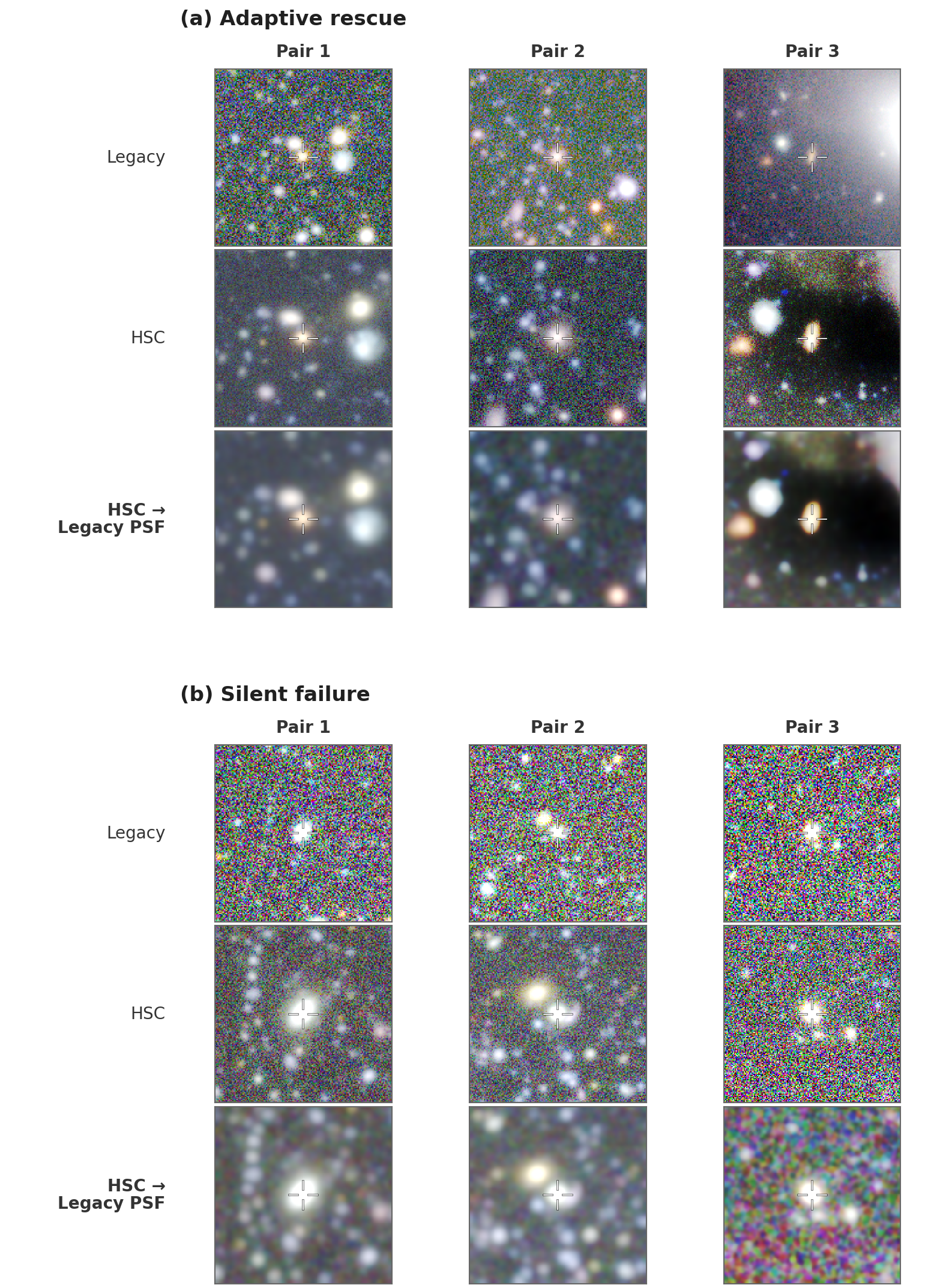}

\caption{
PSF-matching check for representative adaptive-rescue and silent-failure objects. Each column follows the same object from Legacy Survey imaging to the original HSC image and to the HSC image convolved to an approximate Legacy PSF FWHM of $1.2''$. Crosshairs mark the target source. The original and degraded HSC images use the same display normalization.
}
\label{fig:psf_matching}

\end{figure}

Figure~\ref{fig:psf_matching} shows this comparison for representative adaptive rescues and silent failures. Smoothing removes some of the fine structure visible in HSC, but the degraded images do not uniformly reproduce the Legacy appearance. Resolution is therefore one contributor to the cross-survey mismatch, but it does not account for all of the examples
identified by the adaptive analysis.

\small
\end{document}